\documentclass[notitlepage,aps,epsfig,showpacs,floats,twocolumn,amssymb,amsmath,floatfix,
groupedaddress,superscriptaddress]{revtex4-1}
\pdfoutput=1
\usepackage{graphicx} 
\usepackage{dcolumn}
\usepackage{ifpdf}
\usepackage{float} 
\usepackage{amsmath} 
\usepackage{natbib}
\usepackage{bm} 
\usepackage{amsmath,amssymb}
\usepackage{color}
\usepackage{bbm} 
\usepackage{amsfonts}
\usepackage{ulem}
\usepackage{lmodern} 
\usepackage{mathrsfs}
\usepackage[dvipsnames]{xcolor}

\newcommand\bi{\begin{itemize}}
\newcommand\ei{\end{itemize}}

\usepackage[colorlinks=true,citecolor=blue]{hyperref}
\hypersetup{colorlinks=true,citecolor=blue,linkcolor=blue,urlcolor=black}
\begin{document}
\title{Thin active {nematohydrodynamic} layers: asymptotic theories and instabilities}
\author{Mehrana R. Nejad}
\email{mehrana@seas.harvard.edu}
\affiliation{School of Engineering and Applied Sciences, Harvard University, Cambridge, MA 02138, USA}
\author{L. Mahadevan}
\email{lmahadev@g.harvard.edu}
\affiliation{School of Engineering and Applied Sciences, Harvard University, Cambridge, MA 02138, USA}
\affiliation{Departments of Physics, and Organismic and Evolutionary Biology, Harvard University, Cambridge, MA 02138}

\begin{abstract}
Starting from a three-dimensional description of an active nematic layer, we employ an asymptotic theory to derive a series of low-dimensional continuum models 
that capture the coupled dynamics of flat and curved films, including variations in film thickness, shape deformations, internal velocity fields, and the dynamics of orientational order.

Using this asymptotic theory, we  investigate instabilities driven by activity in both the nematic and isotropic phases for cylindrical and flat films. In the flat case, we demonstrate that incorporating shape and thickness variations fundamentally alters the bend and splay nature of instabilities compared to conventional two-dimensional nematic instabilities. In the isotropic phase, we find that both extensile and contractile activity can induce nematic order— in contrast with active nematics on fixed surfaces, where only extensile activity leads to ordering. For the case of curved geometries such as a cylindrical film, we reveal that thickness and shape instabilities are inherently coupled. In the isotropic phase, the emergence of nematic order triggers both thickness and shape instabilities. In the nematic phase, contractile activity induces thickness instabilities, which in turn drive geometric deformations. Our results highlight the crucial interplay between activity, thickness variations, and curvature, providing new insights into the behavior of active nematic films beyond the conventional two-dimensional paradigm that has been studied to date.
\end{abstract}
\maketitle
\section{introduction}
    Hydrodynamic theories of active systems have been instrumental in understanding the dynamics of living matter in both two and three dimensions. Grounded in symmetries and conservation laws, these frameworks have successfully explained a wide range of phenomena observed in biological and synthetic active systems \cite{hatwalne2004rheology,RevModPhys.85.1143,PhysRevLett.125.257801}. Early theoretical descriptions of two-dimensional active matter assumed a strictly 2D environment, employing 2D fields to characterize the dynamics of living systems. These models have provided significant insights into systems confined between rigid boundaries or constrained from deforming in the third dimension, where the shape of the layer remains fixed while active agents move and reorient within the plane \cite{bricard2015emergent,opathalage2019self}.  

In the low-Reynolds-number regime, active nematohydrodynamic theories have successfully explained various phenomena in micron-scale living systems, including molecular motors and active filaments \cite{PhysRevLett.125.218004,decamp2015orientational}, bacterial suspensions \cite{xu2023geometrical,li2019data,yaman2019emergence}, and eukaryotic cells \cite{saw2017topological,saw2018biological}.  
Other recent works have examined instabilities in active nematics confined within a fixed flat layer with fluid interfaces at the top and bottom, revealing instabilities driven by density fluctuations \cite{vskultety2024hydrodynamic}. Additionally, it has been shown that when interfacial number conservation is absent, stable nematic order can emerge at the interface of bulk active nematics, provided the interface remains fixed \cite{maitra2023two}.  

However, many biological systems interact with their surroundings and undergo dynamic shape transformations \cite{meitke,teranishi2023epithelial}. 

Recent studies have begun exploring the interplay between active nematic dynamics and the geometry of the substrate. In systems with fixed curved geometries, it has been shown that curvature plays a crucial role in shaping instabilities within the ordered phase of active systems, often dictating the preferred direction of instability growth \cite{pearce2020defect,alaimo2017curvature,nitschke2018nematic}. Building on this work, other studies have considered the shape deformation of living layers or bulk cylindrical threads, driven by defects or active nematic instabilities, reflecting, for example, the morphological transformations observed in organisms such as \textit{Hydra} \cite{ravichandran2025topology,maroudas2025mechanical,al2023morphodynamics}.

These studies, however, either rely on thin-shell approximations that neglect thickness variations, or assume that the layer does not deform. Yet, both thickness variation and shape deformation are crucial in many key biological processes.
A particularly important example is gastrulation—a critical stage of early embryonic development—during which the three primary germ layers emerge, eventually giving rise to all tissues and organs in the adult organism. During this process, a flat, cylindrical, or spherical sheet undergoes changes in thickness, curvature, or both, driven by cell intercalation and invagination, depending on the species or specific molecular perturbations (see Fig. \ref{fig1}). 

To investigate the role of both shape and thickness variations in the dynamics of active nematic layers, we develop a continuum description using an asymptotic theory that exploits the natural scale separation in these problems—a common approach in low-dimensional descriptions of continuous systems, often referred to as lubrication theory or long-wavelength asymptotic ~\cite{godreche2005hydrodynamics,oron1997long,howell1996models}. Our approach systematically derives the dynamics of a deformable, finite-thickness active nematic layer, where the thickness is small but finite compared to the system's characteristic length. In Sec. \ref{3ddynamicssection}, we introduce the full 3D dynamics of an active nematic layer. Section \ref{highlycurvedfilm} presents the coordinate system and geometrical relations used to describe the curved film, alongside the boundary conditions at the interface. In Sec. \ref{sectionflatlub} and \ref{lubcurv}, we apply the asymptotic theory to derive the effective equations governing the dynamics of the flat and curved layers, respectively. Notably, these two cases involve distinct assumptions in the asymptotic expansion, such that the dynamics of the flat layer cannot simply be recovered as a limiting case of the curved system.

We apply our framework to analyze the stability of both the nematic and isotropic phases in a flat layer (in Sec. \ref{lsafilmthicknessflat}) and a cylindrical geometry (in Sec. \ref{sectionlsa}), comparing our results with instabilities observed in conventional 2D active nematics, which lack thickness and shape variations. Finally, in  Sec. \ref{conclusionsection}, we summarize the results and discuss future directions. As we elaborate, our analysis reveals that deformability and thickness dynamics introduce new instabilities that are absent in traditional 2D active nematics. In particular, in a flat film, allowing for thickness variations modifies the well-known bend (splay) instability, leading to an instability that is not restricted to a specific direction. In the curved case (cylinder), active nematic instabilities drive both thickness and shape deformations. Furthermore, in a curved system with an isotropic ground state, both extensile and contractile activity can induce nematic ordering—contrasting with conventional active nematics, where only extensile activity generates nematic order \cite{santhosh2020activity,PhysRevLett.129.258001}.



\section{3D Dynamics}\label{3ddynamicssection}
\begin{figure*}[t] 
    \centering
    \includegraphics[width=0.9\textwidth]{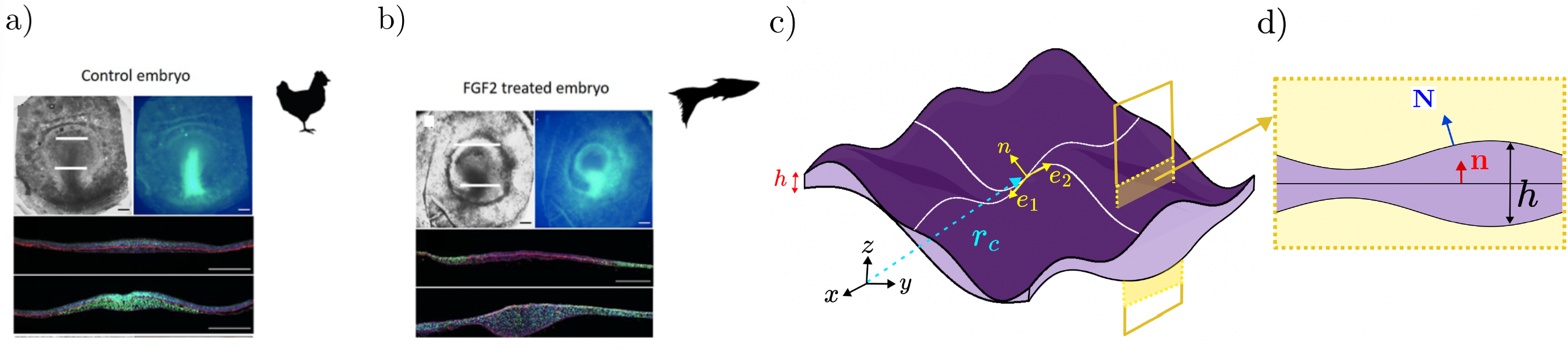}
    \caption{Examples of thickness changes during gastrulation in (a) chick embryo, and (b) chameleon (adapted from Ref.~\cite{chuai2023reconstruction}). (c) Schematic representing the film with finite thickness $h(x_1, x_2, t)$. The film center-surface is represented by $\textbf{r}=\textbf{r}_c$, and the coordinate system $(\textbf{e}_1,\textbf{e}_2,\textbf{n})$ is chosen to be along the direction of the principal curvature of the film center-surface, with the surface normal $\textbf{n}$ locally perpendicular to the center-surface of the film. (d) Schematic representing a cross-section of the film. The unit vector $\textbf{N}(x_1,x_2,t)$ is perpendicular to the film interface. The position of any point in the fluid is given by
$\boldsymbol{r}(x_1, x_2, t) = \boldsymbol{r}_c(x_1, x_2, t) + n\,\mathbf{n}(x_1, x_2, t),$
where $\boldsymbol{r}_c$ is the position of the center surface and $\mathbf{n}(x_1, x_2, t)$ is the unit normal to the surface. The fluid velocity is denoted by $\mathbf{u} = (u(x_1, x_2, t), v(x_1, x_2, t), w(x_1, x_2, t))$, and the velocity of the center surface by $(V_1(x_1, x_2, t), V_2(x_1, x_2, t), V_3(x_1, x_2, t))$. 
We assume that the orientation of the nematic director field lies entirely in the plane of the surface, with its orientation relative to the $\mathbf{e}_1$ axis described by the angle $\theta(x_1, x_2, t)$, and its magnitude described by the scalar order parameter $S(x_1, x_2, t)$. Finally, the curvatures along the $\mathbf{e}_1$ and $\mathbf{e}_2$ directions are denoted by $\kappa_1(x_1, x_2, t)$ and $\kappa_2(x_1, x_2, t)$, respectively.
Scale bars in the top row of panels (a) and (b) represent $500~\mu\mathrm{m}$, and in the bottom row, $250~\mu\mathrm{m}$. }
    \label{fig1}
\end{figure*}

A critical assumption that we make is that it is possible to have an {unambiguous center surface} with respect to which we can define a coordinate system: this is usually possible for embeddings which have one dimension that is small compared to the other two.  Then, the equations governing the dynamics of an active nematic fluid in three dimensions are most naturally expressed in a comoving curvilinear coordinate system with basis vectors as illustrated in Fig.~\ref{fig1}, and denoted by $(\mathbf{e}_1(x_1,x_2,t), \mathbf{e}_2(x_1,x_2,t), \mathbf{n}(x_1,x_2,t))$, which are all functions of the center surface coordinates $(x_1,x_2)$ and time $t$; a precise definition of this orthogonal basis is provided in Sec. ~\ref{highlycurvedfilm}, where we introduce the geometry of the film. Denoting the covariant (contravariant) components of tensors with lower (upper) indices, the dynamics of an active nematic fluid  is governed by the incompressible nematohydrodynamic equations for the fluid velocity $\textbf{u}(x_1,x_2,t) = (u(x_1,x_2,n,t), v(x_1,x_2,n,t), w(x_1,x_2,n,t))$ and the nematic order parameter $\textbf{Q}(x_1,x_2,n,t)$ in 3D:  
\begin{align}
 &\mu \nabla^2 u^i - \nabla^i P + \nabla_k \sigma^{ki}_a = 0,  \label{stokesequa}\\
&\nabla_k u^k = 0, \label{incomp} \\
&\bar{D}_t Q^{ij} =+ \frac{2}{3} \lambda  E^{i j} + Q_{i k} \Omega^{k j} - \Omega^{i}_{\: k} Q^{k j} + \frac{H^{i j}}{\gamma},\label{eqdyq}
\end{align}   
where the active stress is given by $\sigma_{\text{a}}^{ij} = m Q^{ij}$, with $m > 0$ ($m < 0$) corresponding to contractile (extensile) activity, which we assume is large compared to the passive elastic stress. Here $P$ is the pressure which is the Lagrange multiplier enforcing incompressibility,  $\bar{D}_t$, $\boldsymbol{\Omega}$ and $\textbf{E}$ are the time derivative, vorticity and strain rate tensors, in moving coordinates, respectively (see Sec. \ref{dynamicsofq} ), and the molecular field $H^{ij}=-\delta f_Q/\delta Q_{ij}$ governs the relaxation of the nematic tensor towards minimum of the free energy  
\begin{align}
f_Q &= \frac{\mathcal{A}}{4} (1+\mathcal{B}Q^{kl}Q_{kl})^2 + \frac{K_Q}{2} \nabla_m Q^{lk} \nabla^m Q_{lk}, \label{fef}
\end{align}
where we assume a single elastic constant $K_Q$ for simplicity. In the absence of activity, the sign of $\mathcal{B}$ determines whether the system is in the isotropic ($\mathcal{B}>0$) or a nematic ($\mathcal{B}<0$) phase, and Eq.~\eqref{fef} yields the form of the molecular field as
\begin{align}
H^{ij} = -\mathcal{A}(1 + \mathcal{B} Q^{kl}Q_{kl}) \mathcal{B} Q^{ij} + K_Q (\nabla^2 Q)^{ij}.\label{eqmolec}
\end{align}
We note that in a curvilinear coordinate system, the Laplace-Beltrami operator corresponding to the last term has multiple contributions (see SM, Sec. \ref{nematicdiffusion}).

The three-dimensional nature of the problem associated with flow dynamics implies that we must define the nematic order tensor as a rank $3$ tensor. However, here we will assume that the nematic orientation remains within the plane formed by the orthogonal axes $\textbf{e}_1$ and $\textbf{e}_2$, and so we set \( Q_{13} = Q_{23} = 0 \), and \( Q_{33} = -S(x_1, x_2, t) / 3 \). 
Denoting the orientation of the nematic order relative to the $\textbf{e}_1$-axis by $\theta (x_1,x_2,t)$ and the magnitude of the nematic order parameter by $S(x_1,x_2,t)$, it is convenient to represent the nematic order and its orientation field using Pauli matrices $\boldsymbol{\Pi}_{q}$, where $q \in \{1,2,3\}$ (see SM, Sec.~\ref{sthetadynamics}). Specifically, we define 
\begin{align}
\boldsymbol{\Pi}_p &= \sin 2\theta \, \boldsymbol{\Pi}_1 + \cos 2\theta \, \boldsymbol{\Pi}_3, \\
\boldsymbol{\pi} &= \cos 2\theta \, \boldsymbol{\Pi}_1 - \sin 2\theta \, \boldsymbol{\Pi}_3.
\end{align}
Accordingly, the nematic order tensor $\boldsymbol{Q}$ is expressed as 
\begin{align}
\boldsymbol{Q} = \frac{3S}{4} \left( \boldsymbol{\Pi}_p + \boldsymbol{G} \right),
\end{align}
where $G^{ij} =  (\delta^{ij} - 3 \delta^{zi} \delta_z^j)/3$. Utilizing the properties of the Pauli matrices (see SM, Sec. \ref{sthetadynamics}) and using Eq. \eqref{eqdyq}, we can derive the evolution equations for $S$ and $\theta$:  
\begin{align}
&\partial_t S = \frac{2}{3} \operatorname{Tr} \left( \boldsymbol{\Pi}_p \partial_t \boldsymbol{Q} \right), \label{dynamicsofS} \\
&\partial_t \theta = \frac{1}{3 S} \operatorname{Tr} \left( \boldsymbol{\pi} \partial_t \boldsymbol{Q} \right), \label{dynamicsoftheta}
\end{align}  
Equations \eqref{stokesequa}–\eqref{eqdyq} determine the pressure $P$, fluid velocity $\mathbf{u}$, and nematic order $\mathbf{Q}$ once boundary conditions are prescribed. Most previous theories are limited to two dimensions, where the velocity field is a two-dimensional vector field and the nematic tensor is a traceless rank-2 tensor. Here, we deviate from this and now proceed to introduce the geometry of films with finite thickness embedded in three dimensions, with the goal of deriving an asymptotic theory for long wavelength deformations that couple the dynamics of thickness variations to the orientational order and velocity fields.

\section{Active Nematodynamics in $2+\epsilon$ dimensions}\label{highlycurvedfilm}

\subsection{Geometry, Kinetics and Compatibility}
 In terms of the coordinates $\left(x_1, x_2, n\right)$ introduced earlier to describe a general point on the fluid film, the position of any point is given by (see Fig. \ref{fig1})
\[
\boldsymbol{r}(x_1, x_2, t) = \boldsymbol{r}_c\left(x_1, x_2, t\right)+n \textbf{n}(x_1, x_2, t),
\]  
where $\boldsymbol{r}_c$ denotes the film center-surface, and ${\bf n}(x_1,x_2,t)$ is the unit normal to the center surface, with $n$ denoting position in the normal direction. To facilitate the derivation of the model for a general curved film, we choose $x_1$ and $x_2$ such that they parametrize the lines of curvature of the center surface. 

Then deformations of the center surface can be used to define the metric tensor,  
\[
g_{ij} = \frac{\partial \boldsymbol{r}_c}{\partial x_i} \cdot \frac{\partial \boldsymbol{r}_c}{\partial x_j},
\]  
and the curvature tensor,  
\[
B_{ij} = \frac{\partial \boldsymbol{r}_c}{\partial x_i} \cdot \partial_j \textbf{n},
\]  
both of which become diagonal by virtue of the choice of the coordinates.  

Since the sheet is embedded in three dimensions, the fluid sheet thickness $h\left(x_1, x_2, t\right)$ can vary in space and time, so that the two free surfaces of the film, shown in Fig. \ref{fig1}(d) have dynamics. The position of the free surfaces are given by
$\boldsymbol{r}=\boldsymbol{r}_c \pm \frac{1}{2} h \boldsymbol{n} $, where the unit normal to the center-surface (see Fig. \ref{fig1}) is given by 
\begin{align}
\boldsymbol{n}=\left|\frac{\partial \boldsymbol{r}_c}{\partial x_1} \times \frac{\partial \boldsymbol{r}_c}{\partial x_2}\right|^{-1}\left(\frac{\partial \boldsymbol{r}_c}{\partial x_1} \times \frac{\partial \boldsymbol{r}_c}{\partial x_2}\right),
\end{align}
while the orthonormal basis $\left\{\boldsymbol{e}_1, \boldsymbol{e}_2, \boldsymbol{n}\right\}$, illustrated in Fig. \ref{fig2}(a), is defined as
\begin{align}
\boldsymbol{e}_1 = \frac{1}{a_1} \frac{\partial \boldsymbol{r}_c}{\partial x_1}, \quad  
\boldsymbol{e}_2 = \frac{1}{a_2} \frac{\partial \boldsymbol{r}_c}{\partial x_2},
\end{align}
with the coordinate scaling factors given by  
\begin{align}
a_1 = \left|\frac{\partial \boldsymbol{r}_c}{\partial x_1} \right|, \quad  
a_2 = \left|\frac{\partial \boldsymbol{r}_c}{\partial x_2} \right|.
\end{align}
Next, we denote the velocity of the center-surface at $\boldsymbol{r} = \boldsymbol{r}_c$ by  
\begin{equation}
    \frac{\partial \boldsymbol{r}_c}{\partial t} = V_1{(x_1,x_2,t)} \boldsymbol{e}_1 + V_2 {(x_1,x_2,t)} \boldsymbol{e}_2 + V_3 {(x_1,x_2,t)} \boldsymbol{n}.
\end{equation}
Utilizing the orthogonality of the coordinate system and denoting the curvatures along the $x_1$ and $x_2$ directions by $\kappa_1$ and $\kappa_2$, respectively, one can use the spatial derivatives of the basis vectors to show that they must satisfy a set of compatibility relations linking the velocities, curvatures and scaling factors \cite{van1995pressure}   
\begin{align}
   & a_1 \frac{\partial \kappa_1}{\partial x_2} = \left(\kappa_2 - \kappa_1\right) \frac{\partial a_1}{\partial x_2}, \label{relat} \\
   & a_2 \frac{\partial \kappa_2}{\partial x_1} = \left(\kappa_1 - \kappa_2\right) \frac{\partial a_2}{\partial x_1}, \label{relatb} \\
   & \frac{\partial}{\partial x_1} \left(\frac{1}{a_1} \frac{\partial a_2}{\partial x_1} \right) + \frac{\partial}{\partial x_2} \left(\frac{1}{a_2} \frac{\partial a_1}{\partial x_2} \right) + a_1 a_2 \kappa_1 \kappa_2 = 0, \label{relateb}\\
    &V_1 \frac{\partial a_1}{\partial x_2} + V_2 \frac{\partial a_2}{\partial x_1} = a_1 \frac{\partial V_1}{\partial x_2} + a_2 \frac{\partial V_2}{\partial x_1}, \label{gg1} \\
   & a_2 \frac{\partial a_1}{\partial t} = a_2 \frac{\partial V_1}{\partial x_1} + V_2 \frac{\partial a_1}{\partial x_2} - a_1 a_2 \kappa_1 V_3, \label{gg2} \\
   & a_1 \frac{\partial a_2}{\partial t} = a_1 \frac{\partial V_2}{\partial x_2} + V_1 \frac{\partial a_2}{\partial x_1} - a_1 a_2 \kappa_2 V_3. \label{gg3}
\end{align}
As will be shown later, relations \eqref{relat}–\eqref{gg3} will be instrumental in analyzing the behavior of the active film. 
\subsection{Boundary Conditions}
At the free boundaries, given by $n = \pm \frac{1}{2} h(x_1, x_2, t)$, if capillary forces due to surface tension $\Gamma$ are small—i.e., in dimensionless terms, if the Capillary number $Ca= \mu U/\Gamma \gg 1$, which is typically the case in many multicellular tissues—then the stresses vanish on the free surfaces, allowing us to impose zero-stress boundary conditions. This implies that
\begin{align}
\sigma^{ij} N_j = 0.\label{stressfree}
\end{align}
Here $\textbf{N}(x_1,x_2,t)$ denotes the normal to the free surface, which is different from the normal to the center surface $n(x_1,x_2,t)$ but can be expressed in terms of the spatial derivatives of the thickness $h$ (see SM, Sec. \ref{bcsection} and Fig. \ref{fig1}(d)), and $\sigma^{ij}=2\mu E^{ij}-P\delta^{ij}+m Q^{ij}$ is the total stress tensor. In addition, we must also impose a kinematic boundary condition at the free surfaces $n=\pm h/2$, which states that the velocity of the surface is the same as the velocity of the fluid at the interface, i.e.
\begin{align}
\pm \frac{D h}{D t} = w(x_1,x_2,\pm h/2,t).
\end{align}
For a 2-dimensional curved sheet embedded in three dimensions, this kinematic boundary condition can be expressed as  
\begin{align}
    &w - V_3 = 
    \pm \frac{1}{2 l_1} \frac{\partial h}{\partial x_1} [u - V_1 \pm \frac{h}{2} (\kappa_1 V_1 + \frac{1}{a_1} \frac{\partial V_3}{\partial x_1})] \nonumber \\
    &\quad \pm \frac{1}{2 l_2} \frac{\partial h}{\partial x_2} [v - V_2 \pm \frac{h}{2} (\kappa_2 V_2 + \frac{1}{a_2} \frac{\partial V_3}{\partial x_2} ) ]\pm \frac{1}{2} \frac{\partial h}{\partial t} , \label{bckin}
\end{align}
where we have introduced the convective derivative as  
\begin{align}
    \frac{D}{D t} = \frac{\partial}{\partial t} + \left(\boldsymbol{u} - \frac{\partial \boldsymbol{r}_c}{\partial t} - n \frac{\partial \boldsymbol{n}}{\partial t} \right) \cdot \nabla.
\end{align}
In addition, we assume that the nematic tensor remains homogeneous across the thickness, $\textbf{Q}=\textbf{Q}(x_1,x_2,t)$, which is tantamount to stating that the free-surface conditions on the nematic order force it to be parallel to the surface.

Eqs. \eqref{stokesequa}, \eqref{incomp}, \eqref{dynamicsofS}, \eqref{dynamicsoftheta}, and \eqref{relat}–\eqref{gg3} along with the boundary conditions \eqref{stressfree}–\eqref{bckin} form a closed set of equations for the dynamics of the 13 unknowns fields $(u,v,w,P,S,\theta,V_1,V_2,V_3,\kappa_1,\kappa_2,a_2,a_2)$.

\section{Asymptotic theory for weakly curved films}\label{sectionflatlub}
Although the equations describing the dynamics of a thin sheet embedded in three dimensions are complete, they remain somewhat difficult to develop an intuitive understanding of. To do so, we analyze the dynamics for the small transverse deformations of a flat film, to deduce a theory for active nematics that is analogous to that for thin viscous, elastic or viscoelastic films~\cite{howell1996models,teichman2002wrinkling,slim2012buckling,srinivasan2017wrinkling}. 

For small deformations relative  to a flat state, we can represent the center-surface of the film as $\textbf{r}_c = (x_1, x_2, H(x_1, x_2,t))$, and use this small center-surface deformation approximation (Monge Gauge) to define the geometric parameters associated with the film. To incorporate the slenderness of the film with small curvature into our non-dimensionalization of the equations, we let $L$ be a characteristic length scale of the film and $\epsilon L$ a typical thickness, where the dimensionless parameter $\epsilon \ll 1$. We further assume that the transverse deformations are small, i.e., the scaled curvatures $\kappa_i \sim \mathcal{O}(\epsilon)$ and the scaling factors $a_i \sim 1 + \mathcal{O}(\epsilon^2)$, consistent with the Monge Gauge approximation, which we will introduce in Eq. \eqref{curvaturematrix}. We also note that the geometric and kinematic compatibility relations, Eqs.~\eqref{relat}--\eqref{gg3}, become trivial for the nearly flat film. 

Assuming a characteristic velocity scale $\boldsymbol{u} \sim U$, and a characteristic viscosity $\mu \sim \mathscr{M}$, in the limit of small curvature deformations, we now adopt the asymptotic scaling relations for the various fields 
\begin{align*}
    & x_1 = L \tilde{x}_1, \quad x_2 = L \tilde{x}_2, \quad n = \epsilon L \tilde{n}, \\
    & u = U \tilde{u}, \quad v = U \tilde{v}, \quad w = \epsilon U \tilde{w}, \\
    & H = \epsilon L \tilde{H}, \quad \kappa_1 = \epsilon L^{-1} \tilde{\kappa}_1, \quad \kappa_2 = \epsilon L^{-1} \tilde{\kappa}_2, \\
    & h = \epsilon L \tilde{h}, \quad P = \mathscr{M} U L^{-1} \tilde{P}, \quad t = L U^{-1} \tilde{t}, \\
    & \mu = \mathscr{M} \tilde{\mu}.
\end{align*}

Defining the non-unit principal directions of the film as $\boldsymbol{\bar{e}}_1 = a_1 \boldsymbol{e}_1$ and $\boldsymbol{\bar{e}}_2 = a_2 \boldsymbol{e}_2$, we have
\begin{align}
&\boldsymbol{\bar{e}}_{1}=(1, \:0, \: \partial_{1} H), \:\:\: \boldsymbol{\bar{e}}_{2}=(0, 1, \partial_{2} H), \quad \\&\boldsymbol{n}=\frac{\boldsymbol{\bar{e}}_{1} \times \boldsymbol{\bar{e}}_{2}}{\sqrt{g}}=\frac{1}{\sqrt{g}}(-\partial_{1} H,\: -\partial_{2} H,\: 1).
\end{align}
Here, $g$ is the determinant of the metric tensor
$$
\textbf{g}=\boldsymbol{\bar{e}} \boldsymbol{\bar{e}}=\left(\begin{array}{cc}
1+\left(\partial_{2} H\right)^2 & -\partial_{1} H \partial_{2} H \\
-\partial_{1} H \partial_{2} H & 1+\left(\partial_{1} H\right)^2
\end{array}\right) .
$$
For later use, we also note that the curvature tensor is given by
\begin{equation}
\textbf{B}=-\frac{1}{\sqrt{g}}\left(\begin{array}{cc}
\partial_1^2 H & \partial_1 \partial_2 H \\
\partial_1 \partial_2 H& \partial_2^2 H
\end{array}\right).\label{curvaturematrix}
\end{equation}

To move beyond these geometric preliminaries, we apply these scalings to  equations\eqref{stokesequa}, \eqref{incomp}, \eqref{dynamicsofS}, \eqref{dynamicsoftheta}, and \eqref{relat}–\eqref{gg3} and the boundary conditions \eqref{stressfree}–\eqref{bckin},  and seek solutions for the fields $(u,v,w,P,S,\theta,V_1,V_2,V_3,\kappa_1,\kappa_2,a_2,a_2)$  in  asymptotic expansions in powers of the slenderness parameter $\epsilon$ as follows, e.g. 
\begin{align}
   \tilde{u}(x_1,x_2,n,t) \sim u^{(0)}(x_1,x_2,n,t)+\epsilon^2 u^{(2)}(x_1,x_2,n,t)+\ldots.\label{expandelta}
\end{align}
Using the scalings introduced above, we obtain the following estimates for the stress tensor components: $\sigma_{11} \sim \sigma_{22} \sim \sigma_{12} \sim \sigma_{33} \sim \epsilon^0$, and $\sigma_{23} \sim \sigma_{13} \sim \epsilon^{-1}$ (see SM, Sec. \ref{torquebalanceflat}). Applying these scalings, the first-order Stokes equations along $x_1$ and $x_2$ yield (see SM, Sec. \ref{torquebalanceflat})
\begin{equation}
\partial_n^2 u^{(0)} = \partial_n^2 v^{(0)}=0.
\end{equation}
Additionally, the stress-free boundary conditions (Eqs. \eqref{bcs11}-\eqref{bcs22} in SM, Sec. \ref{bcsection}) at leading order give $\partial_n u^{(0)} (\pm h/2)= \partial_n v^{(0)} (\pm h/2)=0$. Consequently, we obtain 
\begin{equation}
v^{(0)}=v^{(0)}(x_1,x_2,t), \:\: u^{(0)}=u^{(0)}(x_1,x_2,t). 
\end{equation}
The leading order incompressibility (Eq. \eqref{incomp}) gives (see SM, Sec. \ref{torquebalanceflat})
\begin{equation}
 \partial_n w^{(0)}= -\partial_1 u^{(0)} -\partial_2 v^{(0)},
\end{equation}
and thus 
\begin{equation}
 w^{(0)}= f(x_1,x_2,t)-(\partial_1 u^{(0)} +\partial_2 v^{(0)}) n.  \label{solinc}
\end{equation}
Combining {Eq. \eqref{solinc}} with the leading order kinematic boundary conditions (Eq. \eqref{bckin}) then leads to 
$f(x_1,x_2,t)=V_3^{(0)}$ and an effective mass conservation in the moving coordinate frame {which} reads
\begin{align}
\partial_t h^{(0)}=&- \partial_1 h^{(0)} \bar{V}_1- \partial_2 h^{(0)} \bar{V}_2-h^{(0)} (\partial_1 u^{(0)} +\partial_2 v^{(0)}),\label{thicknessch}
\end{align}
where $\bar{V}_1 = u^{(0)}-V_1^{(0)}$, and $\bar{V}_2 =v^{(0)}-V_2^{(0)}$, and the last term comes from the $n$ dependence of  $w^{(0)}$. 
In a limiting case where the coordinate system is non-moving, i.e. $V_1^{(0)}=V_2^{(0)}=0$, Eq. \eqref{dynamicsofS} simplifies to 
\begin{align}
&\partial_t h^{(0)}=-u^{(0)} \partial_1 h^{(0)} -v^{(0)} \partial_2 h^{(0)} -h^{(0)} (\partial_1 u^{(0)} + \partial_2 v^{(0)}).\label{massconnx}
\end{align}
\begin{table*}[t]
\centering
\renewcommand{\arraystretch}{1.4}
\begin{tabular}{|c|p{8cm}|}  
\hline
Asymptotic dynamics & Equation \\
\hline
Nematic order & $\partial_t S = \frac{2}{3} \operatorname{Tr} \left( \boldsymbol{\Pi}_p \partial_t \boldsymbol{Q} \right),$ \:\:\: Eq. \eqref{dynamicsofS} \\
\hline
Nematic orientation field & $\partial_t \theta = \frac{1}{3 S} \operatorname{Tr} \left( \boldsymbol{\pi} \partial_t \boldsymbol{Q} \right),$ \: \: \: Eq. \eqref{dynamicsoftheta}\\
\hline
Film thickness & $\partial_t h=- \partial_1 (h u)- \partial_2 (h v),$ \:\:\: Eq. \eqref{thicknessch} \\
\hline
In-plane Stokes equation & 
\begin{minipage}[t]{6cm}
$\partial_1 T_{11} + \partial_2 T_{12} = 0, \:\:\:\:\:\: \:\:\:\:\:\:\:\:\:\:\:\:\:\:\: \:\:\:\:\:\:\:\:\:\:\:\:\:\:\:\:\:\:\:\:\:\: \:\:\:\:\:\:\:\:\:\:\:\:\:\:\:\:\:$\\
$\partial_1 T_{12} + \partial_2 T_{22} = 0,\:\:\:\:\:\:\:\:\:\:\:\: \:\:\:\:\:\:\:\:\:\:\:\:\:\:\:\:\:\:\:\:\:\:\:\: \:\:\:\:\:\:\:\:\:\:\:\:\:\:\:\:\:\:\:\:\:\:\:\:$ \\
$T_{11} = m h (Q_{11} - Q_{33}) + 2\mu h\, \partial_2 v + 4\mu h\, \partial_1 u,$ \\
$T_{22} = m h (Q_{22} - Q_{33}) + 2\mu h\, \partial_1 u + 4\mu h\, \partial_2 v,$ \\
$T_{12} = m h Q_{12} + \mu h\, \partial_2 u + \mu h\, \partial_1 v.$  $ \:\:\:\:\:\: \:\:\:\:\:\:\:\:\:\:\:\:\:\:$
 \\
 \: \: \: Eqs. \eqref{filmforcebal1}-\eqref{eqwar}\\
\end{minipage} \\
\hline
\end{tabular}
\caption{\textbf{Effective equations governing the dynamics of a flat film with a finite thickness.} The set of effective equations describes the dynamics of the nematic orientation field $\theta$, the magnitude of the nematic order $S$, the film thickness $h$, and the fluid flow along $\mathbf{e}_1$ and $\mathbf{e}_2$ {directions} (denoted by $u$ and $v$, respectively).
}\label{tableflateq}
\end{table*}
The first two terms on the right-hand side of Eq.~\eqref{massconnx} represent the advection of thickness inhomogeneities by the flows, while the last two terms show that fluid flows with positive divergence \((\partial_1 u^{(0)} + \partial_2 v^{(0)} > 0)\) increase the thickness, and those with negative divergence \((\partial_1 u^{(0)} + \partial_2 v^{(0)} < 0)\) decrease the thickness.
For the pressure, the leading-order nonzero transverse Stokes equation (Eq. \eqref{sadsadg1} in SM, Sec. \eqref{stressexpansion}) yields
\begin{align}
\partial_n P^{(0)} =\mu (2 \partial_n^2 w^{(0)} +\partial_n \partial_2 v^{(0)} +\partial_n \partial_1 u^{(0)})=0,\label{pderiv}
\end{align}
 and the stress-free boundary condition (Eq. \eqref{bcs33} in SM, Sec. \eqref{bcsection}) reads
 \begin{align}
P^{(0)}(\pm h/2) =s m Q_{33} -2 \mu (\partial_1 u^{(0)} +\partial_2 v^{(0)}).\label{bcpk}
\end{align}
Eqs. \eqref{bcpk} and \eqref{pderiv} then give the solution for the pressure, which reads 
 \begin{align}
P^{(0)}(x_1,x_2,t) = m Q_{33} -2 \mu (\partial_1 u^{(0)} +\partial_2 v^{(0)}).\label{pressureflat}
\end{align}
Given the pressure, we then integrate the in-plane Stokes equations (Eqs. \eqref{sadsad1} and \eqref{sadsad2}) across the thickness, retaining the first-order non-zero terms, and subsequently obtain the effective in-plane Stokes equation. Omitting the superscript
$^{(0)}$ on the fields, the effective in-plane Stokes equation then reads  
 \begin{align}
&\partial_1 T_{11}+\partial_2 T_{12}=0,\label{filmforcebal1}\\&
\partial_1 T_{12}+\partial_2 T_{22}=0,\label{filmforcebal2}    
\end{align}
where the components of the symmetric tension tensor $\textbf{T}$ are defined as
\begin{align}
    & T_{11}=m h (Q_{11}-Q_{33}) +2 \mu h \partial_2 v+4 \mu h \partial_1 u,\\
    & T_{22}=m h (Q_{22}-Q_{33}) +2 \mu h \partial_1 u+4 \mu h \partial_2 v,\\
     & T_{12}= m h Q_{12} + \mu h \partial_2 u+ \mu h \partial_1 v. \label{eqwar}
\end{align}
In the absence of activity ($m=0$), Eqs. \eqref{filmforcebal1} and \eqref{filmforcebal2} simplify to the effective Stokes equation for thin viscous films, used to describe the 
float glass process \cite{van1995pressure,narayanaswamy1977one}. In extensile (contractile) systems, activity leads to extensional (compressional) stress in the direction of the nematic order. In addition to the Stokes equation, having the first order non-zero stress components, we use the torque balance on the film elements to find the dynamics of the center-surface. This reads (see SM, Sec. \ref{torquebalanceflat} for details)
\begin{equation}
      T_{11} \partial_1^2 H  +2 T_{12} \partial_1 \partial_2 H  +T_{22} \partial_2^2 H =0.\label{torquebalance}
\end{equation}
{Eq. \eqref{torquebalance} is an} anisotropic generalization of the Laplace equation for the center-surface, {and} is mathematically well-posed only when the ellipticity condition {$ T_{22} T_{11} -T_{12}^2 > 0$} holds. This physically corresponds to the condition that the center-surface of a flat film does not buckle when activity is uniformly distributed across the thickness. If this condition is not met, e.g. for strongly contractile active stresses, one has to revisit the asymptotic scaling assumptions underlying the theory, e.g. making the out-of-plane velocity large compared to the {in-plane velocity~\cite{buckmaster1975buckling,howell1996models}}.

Then Eqs.~\eqref{dynamicsofS}, \eqref{dynamicsoftheta}, \eqref{thicknessch}, \eqref{torquebalance}, and \eqref{filmforcebal1}--\eqref{filmforcebal2} form a closed set of equations governing the dynamics of the fields $S$, $\theta$, $h$, $u$, and $v$ in terms of $(x_1, x_2, t)$ in the approximately flat case, thus providing a complete description of the living film. These are summarised in Table \ref{tableflateq}.  
We utilize these equations in Sec.~\ref{lsafilmthicknessflat} to investigate the linear stability of an active film in both the nematic and isotropic phases.
We note that 2D descriptions of active nematics typically ignore both the dynamics of the thickness (Eq.~\eqref{massconnx}) and the evolution of the center-surface shape (Eq.~\eqref{torquebalance}). If we impose a constant thickness, Eq.~\eqref{massconnx} reduces to the 2D incompressibility condition, $\partial_1 u + \partial_2 v = 0$. Fixing the shape of the monolayer in our framework corresponds to assuming $H$ is constant in Eq.~\eqref{torquebalance}, thereby allowing us to ignore this equation. 
However, even under these constraints, our formulation of the Stokes flow in Eqs.~\eqref{filmforcebal1}--\eqref{filmforcebal2} differs from a purely 2D description of active nematics. Notably, it retains the third component of the nematic tensor ($Q_{33}$) and includes an effective pressure term.
\subsection{Linear stability analysis of flat film}\label{lsafilmthicknessflat}
In the following subsections, we perform a linear stability analysis of both the nematic and isotropic phases in flat films allowing for both thickness and shape variations. The magnitude of the nematic order is determined by the free energy parameter $\mathcal{B}$ (see Eq. \eqref{fef}).
Specifically, we investigate the instability of the nematic phase and the isotropic phase by selecting $\mathcal{B}=-2/3$ for the ordered nematic phase
($S=1$) and $\mathcal{B}=2/3$ for the 
isotropic phase ($S=0$).
 Additionally, we consider $\lambda\geq0$, which corresponds to systems with particles with an elongated shape.
Without loss of generality, we assume that the director is aligned along the $\textbf{e}_1$-axis. Employing the Fourier representation $\delta \tilde{f}(\textbf{q},\omega) = \int d\textbf{r} \, f(x,y) \, e^{-i\textbf{q} \cdot \textbf{r} + \omega t}$, where the wavevector is given by $\textbf{q} = q (\cos \phi, \sin \phi)$, for the perturbations in the fields $u,v,S,\theta,h$ 
we then use Eqs. \eqref{filmforcebal1}-\eqref{filmforcebal2} to solve for the steady-state values of the flow fields and their perturbations $(\delta u, \delta v)$, which are then substituted into the dynamics of the nematic order parameter (Eqs. \eqref{dynamicsofS}-\eqref{dynamicsoftheta}), and thickness (Eq. \eqref{massconnx}), to find the dynamics of the perturbations in the orientation field $\delta \theta$, magnitude of the nematic order $\delta S$, and film thickness $\delta h$.
The onset of instability is characterized by the condition $\Re (\omega)>0$.
 \subsubsection{Nematic Phase}
Using Eqs.~\eqref{dynamicsofS}--\eqref{dynamicsoftheta}, \eqref{massconnx}, and \eqref{filmforcebal1}--\eqref{filmforcebal2} we can study the linear stability of a phase characterized by the steady-state solutions $u_s = v_s = \theta_s = 0$, $h_s=h_0$ and $S_s = 1$, corresponding to a nematic phase aligned along the $\boldsymbol{e}_1$ 
  \begin{figure}[h] 
    \centering
    \includegraphics[width=0.45\textwidth]{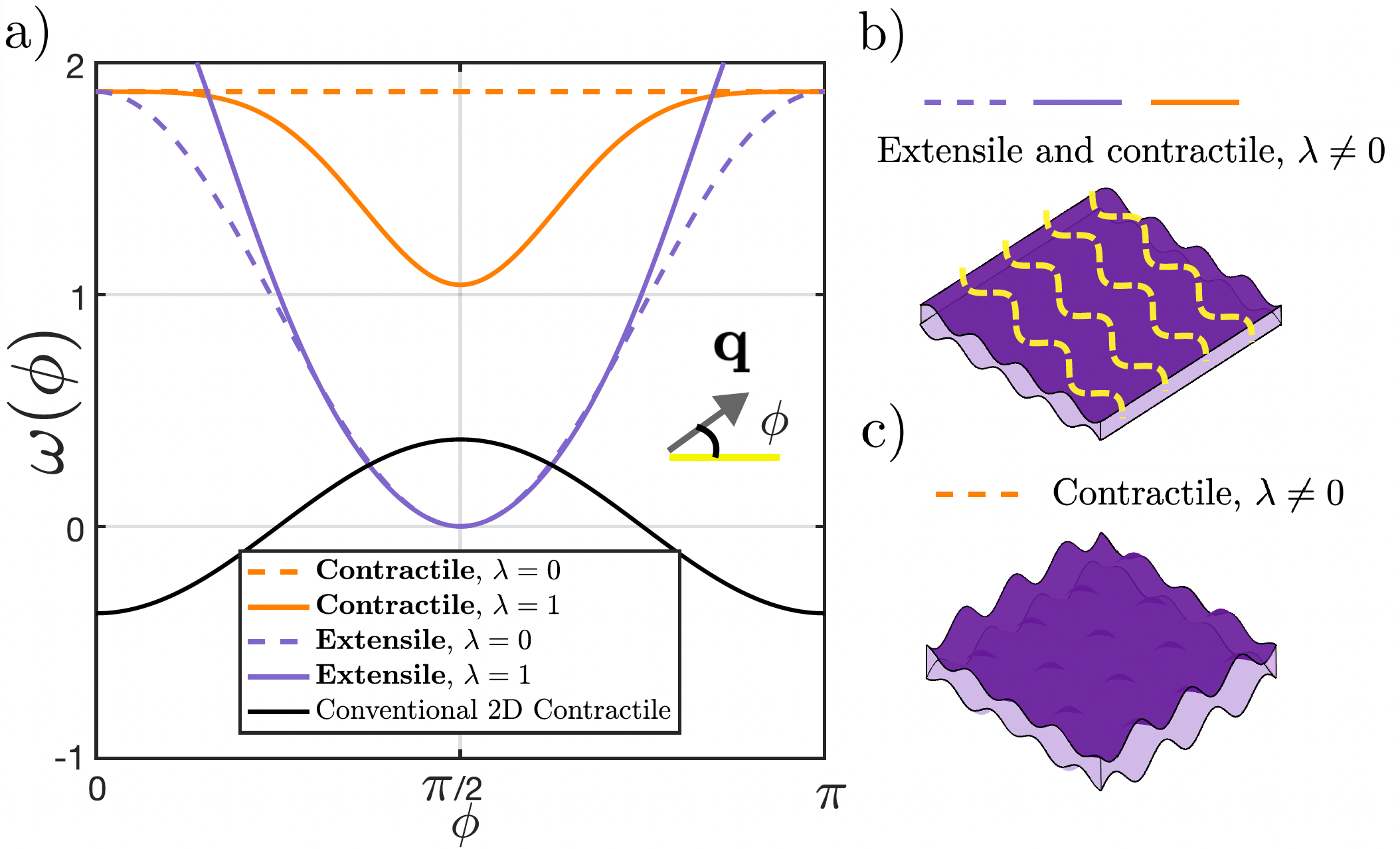}
    \caption{\textbf{Instability of the nematic phase of the living film with planar geometry.} (a) Dispersion relation (Eq. \eqref{fullw}) for contractile (orange) and extensile (purple) systems, with and without flow alignment $\lambda$. Solid (dashed) line   corresponds to $\lambda \neq 0$ ($\lambda = 0$). \(\omega > 0\) indicates the onset of the instability, and $\phi$   denotes the angle between the perturbation wave vector $\textbf{q}$ shown as a grey arrow with respect to the direction of the nematic order (shown in yellow). (b) In the presence of flow alignment, nematic order and constant thickness are most unstable for perturbations along the direction of the order (\(\phi = 0\)) in both contractile and extensile systems. This contrasts with 2D active nematics, which exhibit bend (splay) instability in extensile (contractile) systems. (c) Contractile systems without flow alignment (Eq. \eqref{eqwc}) show an instability that occurs in all directions and is independent of the perturbation angle $\phi$.}
    \label{fig3}
\end{figure}
axis with no flow. For a non-zero flow aligning parameter, the growthrate of the thickness and the nematic order are coupled (see SM, Sec. \ref{nematicflat}). As shown in Fig. \ref{fig3}(a), in the presence of the flow alignment, bend perturbations ($\phi=0,\pi$) have the maximum growth rate for both extensile and contractile systems. 

This highlights a significant distinction from conventional 2D active nematics, where only bend (splay) perturbations grow in extensile (contractile) systems \cite{PhysRevLett.89.058101}. In the limit of a zero flow-alignment parameter ($\lambda=0$), the leading-order growth rate of perturbations in the wave vector simplifies—respectively for contractile and extensile systems—to the growth rates $\omega_c$ and $\omega_e$, given by (see SM Sec. \ref{nematicflat}, Eq. \eqref{fullw})
\begin{align}
&\omega_c=\frac{3 m }{8 \mu}-\frac{3 K_Q q^2 \sin^2\phi}{\gamma (3+\cos 2\phi)}, \label{eqwc}\\&
\omega_e=-\frac{3 m }{8 \mu} \cos^2 \phi -\frac{6 K_Q q^2 \cos^2\phi}{\gamma (3+\cos 2\phi)}. \label{eqwe}
\end{align}
Eq. \eqref{eqwc} shows that contractile activity results in a uniform growth rate for all perturbations, irrespective of their direction (see Fig. \ref{fig3}(a)). Including higher-order terms (see SM Sec. \ref{nematicflat}, Eq. \eqref{fullw}) in the wavevector shows that, similar to 2D active nematics, the length scale over which the system stabilizes is set by the active length scale
 $\ell_a=(K_Q \mu/|m \gamma|)^{1/2}$. 
In the limit where $\delta h=0$, the growth rate reduces to that of conventional 2D active nematics \cite{PhysRevLett.89.058101}:
\begin{align}
\omega=-\frac{m (\lambda  (3 \cos (4 \phi)+5)+18 \cos (2 \phi))}{48 \mu }.
\end{align}
 \subsubsection{Isotropic Phase}
 Using Eqs.~\eqref{dynamicsofS}--\eqref{dynamicsoftheta}, \eqref{massconnx}, and \eqref{filmforcebal1}--\eqref{filmforcebal2} we can study the linear stability of a phase characterized by the steady-state solutions $u_s = v_s = \theta_s = 0$, $h_s=h_0$ and $S_s = 0$, corresponding to an isotropic phase with the director aligned along the $x_1$ axis with no flow. 
Substituting the fourier decomposition $\delta \tilde{f}(\textbf{q},\omega)$ for the fields $(S,\theta,h,u,v)$, the growth rate of the order parameter is given by
\begin{align}
\omega = -\frac{m \lambda }{6 \mu}\cos^2 \phi \left( \cos^2 \phi+7 \sin^2 \phi\right)-\gamma^{-1}(\mathcal{A} \mathcal{B} +3 K_Q q^2),
\end{align}
and the dynamics of the thickness are coupled to the dynamics of the nematic order through activity, and reads
\begin{align}
 \delta h \omega=3 m h_0 \cos^2 \phi \delta S/(8 \mu).
\end{align}
The magnitude of the order grows similarly to that observed in normal 2D active nematics \cite{santhosh2020activity}. The key difference is that, as the nematic order forms and perturbations in $\delta S$ grows, perturbations in $\delta h$ also grow, and thickness modulations appear in the system.
\section{Asymptotic theory for curved thin film}\label{lubcurv}
We now incorporate the slenderness of the film into our nondimensionalization of the equations for a general curved film. Let $L$ be a characteristic length scale of the film and $\epsilon L$ a typical thickness, where the dimensionless parameter $\epsilon \ll 1$. We assume a characteristic velocity scale $\boldsymbol{u} \sim U$, a typical curvature of order $\kappa \sim 1/L$, and a characteristic viscosity $\mu \sim \mathscr{M}$. Without loss of generality, we take the coordinates $x_1$ and $x_2$ to have dimensions of length. Accordingly, we introduce the following nondimensional scalings:
\begin{align*}
    & x_1 = L \tilde{x}_1, \quad x_2 = L \tilde{x}_2, \quad n = \epsilon L \tilde{n}, \\
    & u = U \tilde{u}, \quad v = U \tilde{v}, \quad w = U \tilde{w}, \\
    & \boldsymbol{r}_c = L \tilde{\boldsymbol{r}}_c, \quad \kappa_1 = L^{-1} \tilde{\kappa}_1, \quad \kappa_2 = L^{-1} \tilde{\kappa}_2, \\
    & h = \epsilon L \tilde{h}, \quad P = \mathscr{M} U L^{-1} \tilde{P}, \quad t = L U^{-1} \tilde{t}, \\
    & \mu = \mathscr{M} \tilde{\mu}.
\end{align*}
Similar to the flat film, we assume that the nematic tensor remains homogeneous across the thickness and apply these scalings to nondimensionalize the Stokes equations (Eqs. \eqref{stokesequa} and \eqref{incomp}) along with the corresponding boundary conditions. We would like to point out that in contrast to the flat film, where the curvature was of the order of $\mathcal{O}(\epsilon)$, here the curvature is of the order of $\mathcal{O}(\epsilon^0)$.
Similar to the flat film, solutions of equations \eqref{stokesequa}, \eqref{incomp}, \eqref{dynamicsofS}, \eqref{dynamicsoftheta}, and \eqref{relat}–\eqref{gg3} and the boundary conditions \eqref{stressfree}–\eqref{bckin} for the fields $(u,v,w,P,S,\theta,V_1,V_2,V_3,\kappa_1,\kappa_2,a_2,a_2)$  are subsequently sought in the form of asymptotic expansions in powers of the slenderness parameter $\epsilon$ as follows, e.g. 
\begin{align}
   \tilde{u} (x_1,x_2,n,t) \sim u^{(0)}(x_1,x_2,n,t)+\epsilon u^{(1)}(x_1,x_2,n,t)+\ldots.\label{expandelta}
\end{align}
Using the scalings introduced above and the definition of the stress tensor (see SM, Sec. \ref{forcebalanceeffective}), we obtain the following scaling estimates:  
\begin{align*}
    \sigma_{22}, \sigma_{11}, \sigma_{12} &\sim \mathcal{O}(\epsilon^0), \\
    \sigma_{33}, \sigma_{13}, \sigma_{32} &\sim \mathcal{O}(\epsilon^{-1}).
\end{align*}
Using these scalings, the leading-order Stokes equation (Eq.~\eqref{stokesequa}) reduces to
\begin{align}
    \partial_{{n}}^2 {u}^{(0)} = \partial_{{n}}^2 {v}^{(0)} = \partial_{{n}}^2 {w}^{(0)} = 0, \label{stok0}
\end{align}
and the incompressibility condition (Eq. \eqref{incomp}) simplifies to  
\begin{align}
    \partial_{{n}} {w}^{(0)} = 0. \label{incompcurv0}
\end{align}
Meanwhile, the leading-order stress free boundary conditions at $n = \pm \frac{h}{2}(x_1, x_2, t)$ are (see Eqs. \eqref{bcs11}–\eqref{bcs33} in SM)
\begin{align}
    \partial_{n} {u}^{(0)} = \partial_{n} {v}^{(0)} = \partial_{n} {w}^{(0)} = 0. \label{bczero}
\end{align}
{Using \eqref{bczero},} the kinematic boundary condition (Eq. \eqref{bckin}) at $n = \pm h/2$ simplifies to  
\begin{align}
    {w}^{(0)} = V_3(x_1,x_2,t). \label{kinematicbc}
\end{align}
Using Eqs. \eqref{stok0}, \eqref{incompcurv0}, and \eqref{kinematicbc}, we find that the first order velocity components are independent of $n$, yielding  
\begin{align}
    {u}^{(0)} = {u}^{(0)} (x_1, x_2, t), \quad {v}^{(0)} = {v}^{(0)} (x_1, x_2, t), \quad {w}^{(0)} = V_3^{(0)}. \label{solcurv0}
\end{align}
 For clarity and conciseness, we omit the superscript
$^{(0)}$ on the fields in the remainder of the manuscript. We retain the indices for higher-order fields in $\epsilon$ to avoid confusion. \\
Using the identities \eqref{gg2} and \eqref{gg3}, one can then find an equation representing the conservation of mass, which reads (see SM, Sec. \ref{forcebalanceeffective})
\begin{align}
&\partial_t\left(a_1 a_2 h\right)+\partial_1\left(a_2 h\bar{V}_1\right)+\partial_2\left(a_1 h\bar{V}_2\right)=0,\label{masscon}
\end{align}
where we have defined $\bar{V}_1 = u^{(0)}-V_1^{(0)}$, and $\bar{V}_2 =v^{(0)}-V_2^{(0)}$. For a non-moving film ($\partial_t a_1=\partial_t a_2=V_1=V_2=0$) with a symmetric shape ($\partial_i a_j=0$ for any $i,j \in \{1,2\}$), Eq. \eqref{masscon} simplifies to
\begin{align}
&\partial_t h=-u^i \nabla_i h  -h \nabla_i u^i,\label{massconn}
\end{align}
 which is equivalent to the mass conservation  {in} Eq. \eqref{massconnx} but now for a curved film.
The force balance perpendicular to the film (Eq. \eqref{stokesequa}), with the stress free boundary condition (Eq. \eqref{stressfree}) gives the zero order pressure. Using the stress-free boundary conditions and Stokes equation along the $\textbf{e}_1$ and $\textbf{e}_2$ directions, the first-order in-plane fluid velocities then read (see SM, Sec. \ref{forcebalanceeffective})
\begin{align}
&u^{(1)}=F_1\left(x_1, x_2, t\right)-n\left(\kappa_1 {u}+\frac{1}{a_1} \partial_1 V_3\right)\nonumber \\& -\frac{1}{\mu}\int^n \: Q_{13}(x_1,x_2,n',t) \: m(x_1,x_2,n',t) dn' ,\label{vel1delta}\\
&v^{(1)}=F_2\left(x_1, x_2, t\right)-n\left(\kappa_2 {v}+\frac{1}{a_2}\partial_2 V_3\right)\nonumber \\&-\frac{1}{\mu}\int^n \: Q_{23}(x_1,x_2,n',t) \: m(x_1,x_2,n',t) dn' ,\label{vel2delta}
\end{align}
\begin{table*}[t]
\centering
\renewcommand{\arraystretch}{1.4}
\begin{tabular}{|c|p{9cm}|}  
\hline
Asymptotic dynamics & Equation \\
\hline
Nematic order & $\partial_t S = \frac{2}{3} \operatorname{Tr} \left( \boldsymbol{\Pi}_p \partial_t \boldsymbol{Q} \right),$   \:\:\: Eq. \eqref{dynamicsofS} \\
\hline
Nematic orientation field & $\partial_t \theta = \frac{1}{3 S} \operatorname{Tr} \left( \boldsymbol{\pi} \partial_t \boldsymbol{Q} \right),$  \:\:\:\: Eq. \eqref{dynamicsoftheta} \\
\hline
Film thickness & $\partial_t\left(a_1 a_2 h\right)+\partial_1\left(a_2 h\bar{V}_1\right)+\partial_2\left(a_1 h\bar{V}_2\right)=0,$   \:\:\: Eq. \eqref{masscon}\\
\hline
Compatibility relations & 
\begin{minipage}[t]{8cm}
$a_1 \frac{\partial \kappa_1}{\partial x_2} = \left(\kappa_2 - \kappa_1\right) \frac{\partial a_1}{\partial x_2} \:\:\:\:\:\:\:\:\:\:\: \:\:\:\:\:\:\:\:\:\:\:\:\:\:\:\:\:\:\:\:\:\:\:\:\:\:\:\:\:\:\:\:\:\:\:\:\:\:\:\:\:\:\:\:\:\:\:\:\:\:\:\:\:\:\:\:\:\:\:\:\:\:\:\:\:\:\:\: $\\
$a_2 \frac{\partial \kappa_2}{\partial x_1} = \left(\kappa_1 - \kappa_2\right) \frac{\partial a_2}{\partial x_1},\:\:\:\:\:\:\:\:\:\:\: \:\:\:\:\:\:\:\:\:\:\:\:\:\:\:\:\:\:\:\:\:\:\:\:\:\:\:\:\:\:\:\:\:\:\:\:\:\:\:\:\:\:\:\:\:\:\:\:\:\:\:\:\:\:\:\:\:\:\:\:\:\:\:\:\:\:\:\:$ \\
$\frac{\partial}{\partial x_1} \left(\frac{1}{a_1} \frac{\partial a_2}{\partial x_1} \right) + \frac{\partial}{\partial x_2} \left(\frac{1}{a_2} \frac{\partial a_1}{\partial x_2} \right) + a_1 a_2 \kappa_1 \kappa_2 = 0,\:\:\:\:\:\:\:\:\:\:\: \:\:\:\:\:\:\:\:\:\:\:\:\:\:\:\:\:\:\:\:\:\:\:\:\:\:\:\:\:\:\:\:\:\:\:\:\:\:\:\:\:\:\:\:\:\:\:\:\:\:\:\:\:\:\:\:\:\:\:\:\:\:\:\:\:\:\:\:$ 
 Eqs. \eqref{relat}-\eqref{relateb}
\\
\end{minipage} \\
\hline
Dynamic relations & 
\begin{minipage}[t]{8cm}
$V_1 \frac{\partial a_1}{\partial x_2} + V_2 \frac{\partial a_2}{\partial x_1} = a_1 \frac{\partial V_1}{\partial x_2} + a_2 \frac{\partial V_2}{\partial x_1}\:\:\:\:\:\:\:\:\:\:\: \:\:\:\:\:\:\:\:\:\:\:\:\:\:\:\:\:\:\:\:\:\:\:\:\:\:\:\:\:\:\:\:\:\:\:\:\:\:\:\:\:\:\:\:\:\:\:\:\:\:\:\:\:\:\:\:\:\:\:\:\:\:\:\:\:\:\:\:$\\
$a_2 \frac{\partial a_1}{\partial t} = a_2 \frac{\partial V_1}{\partial x_1} + V_2 \frac{\partial a_1}{\partial x_2} - a_1 a_2 \kappa_1 V_3,\:\:\:\:\:\:\:\:\:\:\: \:\:\:\:\:\:\:\:\:\:\:\:\:\:\:\:\:\:\:\:\:\:\:\:\:\:\:\:\:\:\:\:\:\:\:\:\:\:\:\:\:\:\:\:\:\:\:\:\:\:\:\:\:\:\:\:\:\:\:\:\:\:\:\:\:\:\:\:$ \\
$ a_1 \frac{\partial a_2}{\partial t} = a_1 \frac{\partial V_2}{\partial x_2} + V_1 \frac{\partial a_2}{\partial x_1} - a_1 a_2 \kappa_2 V_3,\:\:\:\:\:\:\:\:\:\:\: \:\:\:\:\:\:\:\:\:\:\:\:\:\:\:\:\:\:\:\:\:\:\:\:\:\:\:\:\:\:\:\:\:\:\:\:\:\:\:\:\:\:\:\:\:\:\:\:\:\:\:\:\:\:\:\:\:\:\:\:\:\:\:\:\:\:\:\:$ Eqs. \eqref{gg1}-\eqref{gg3}
 \\
\end{minipage} \\
\hline
Force balance perpendicular to the film & 
\begin{minipage}[t]{8cm}
$(2 \kappa_1+\kappa_2)(a_2 \partial_t a_1+\bar{V}_2 \partial_2 a_1+a_2 \partial_1 \bar{V}_1)
\:\:\:\:\:\:\:\:\:\:\: \:\:\:\:\:\:\:\:\:\:\:\:\:\:\:\:\:\:\:\:\:\:\:\:\:\:\:\:\:\:\:\:\:\:\:\:\:\:\:\:\:\:\:\:\:\:\:\:\:\:\:\:\:\:\:\:\:\:\:\:\:\:\:\:\:\:\:\:$\\
$+\left(\kappa_1+2 \kappa_2\right)\left(a_1 \partial_t a_2+\bar{V}_1 \partial_1 a_2+a_1 \partial_2 \bar{V}_2\right)\:\:\:\:\:\:\:\:\:\:\: \:\:\:\:\:\:\:\:\:\:\:\:\:\:\:\:\:\:\:\:\:\:\:\:\:\:\:\:\:\:\:\:\:\:\:\:\:\:\:\:\:\:\:\:\:\:\:\:\:\:\:\:\:\:\:\:\:\:\:\:\:\:\:\:\:\:\:\:$ \\
$ +a_1 a_2   m \left(\kappa_1 (Q_{11}-Q_{33})+\kappa_2 (Q_{22}-Q_{33})\right)/(2 \mu ) =0.\:\:\:\:\:\:\:\:\:\:\: \:\:\:\:\:\:\:\:\:\:\:\:\:\:\:\:\:\:\:\:\:\:\:\:\:\:\:\:\:\:\:\:\:\:\:\:\:\:\:\:\:\:\:\:\:\:\:\:\:\:\:\:\:\:\:\:\:\:\:\:\:\:\:\:\:\:\:\:$ Eqs. \eqref{forcebalancentop}
 \\ 
\end{minipage} \\
\hline
Force balance parallel to the film along $\textbf{e}_{1/2}$  direction& 
\begin{minipage}[t]{8cm}
$\partial_{1/2} \{\frac{2\mu h }{a_{1/2} } [2\left(a_{2/1} \partial_t a_{1/2} +\bar{V}_{2/1} \partial_{2/1} a_{1/2} +a_{2/1} \partial_{1/2}\bar{V}_{1/2}\right) \:\:\:\:\:\:\:\:\:\:\: \:\:\:\:\:\:\:\:\:\:\:\:\:\:\:\:\:\:\:\:\:\:\:\:\:\:\:\:\:\:\:\:\:\:\:\:\:\:\:\:\:\:\:\:\:\:\:\:\:\:\:\:\:\:\:\:\:\:\:\:\:\:\:\:\:\:\:\:$\\
$+\left(a_{1/2} \partial_t a_{2/1}+\bar{V}_{1/2} \partial_{1/2} a_{2/1} +a_{1/2}  \partial_{2/1}\bar{V}_{2/1}\right)]\}
  +F_{1/2}^a\:\:\:\:\:\:\:\:\:\:\: \:\:\:\:\:\:\:\:\:\:\:\:\:\:\:\:\:\:\:\:\:\:\:\:\:\:\:\:\:\:\:\:\:\:\:\:\:\:\:\:\:\:\:\:\:\:\:\:\:\:\:\:\:\:\:\:\:\:\:\:\:\:\:\:\:\:\:\:$ \\
$+\frac{1}{a_{1/2}} \partial_{2/1}\{\frac{\mu a_{1/2}  h }{a_{2/1} }\left[a_{1/2} \partial_{2/1}\bar{V}_{1/2}+a_{2/1} \partial_{1/2}\bar{V}_{2/1}\right]\}=\:\:\:\:\:\:\:\:\:\:\: \:\:\:\:\:\:\:\:\:\:\:\:\:\:\:\:\:\:\:\:\:\:\:\:\:\:\:\:\:\:\:\:\:\:\:\:\:\:\:\:\:\:\:\:\:\:\:\:\:\:\:\:\:\:\:\:\:\:\:\:\:\:\:$ 
 \\
$ \frac{2 \mu h }{a_{1/2}  a_{2/1} } \partial_{1/2} a_{2/1} \left[\left(a_{2/1} \partial_t a_{1/2} +\bar{V}_{2/1}\partial_{2/1} a_{1/2} +a_{2/1}  \partial_{1/2}\bar{V}_{1/2}\right)\right]\:\:\:\:\:\:\:\:\:\:\: \:\:\:\:\:\:\:\:\:\:\:\:\:\:\:\:\:\:\:\:\:\:\:\:\:\:\:\:\:\:\:\:\:\:\:\:\:\:\:\:\:\:\:\:\:\:\:\:\:\:\:\:\:\:\:\:\:\:\:\:\:$ 
 \\
 $ +2\left(a_{1/2} \partial_t a_{2/1} +\bar{V}_{1/2} \partial_{1/2} a_{2/1}+a_{1/2} \partial_{2/1}\bar{V}_{2/1} \right) ],\:\:\:\:\:\:\:\:\:\:\: \:\:\:\:\:\:\:\:\:\:\:\:\:\:\:\:\:\:\:\:\:\:\:\:\:\:\:\:\:\:\:\:\:\:\:\:\:\:\:\:\:\:\:\:\:\:\:\:\:\:\:\:\:\:\:\:\:\:\:\:\:\:\:\:\:\:\:\:$ Eqs. \eqref{longforcebalance} and \eqref{secondfbeq}
 \\
\end{minipage} \\
\hline
\end{tabular}
\caption{\textbf{Effective equations governing the dynamics of a general curved film with a finite thickness.} The set of 12 equations describes the dynamics of the nematic orientation field $\theta$, the magnitude of the nematic order $S$, the film thickness $h$, and the fluid flow along $\mathbf{e}_1$ and $\mathbf{e}_2$ directions (denoted by $u$ and $v$, respectively), and the shape of the film (through the fields $V_3$, $V_i$, $\kappa_i$, $a_i$ for $i \in \{1,2\}$). The relative velocities $\bar{V}_1$ and $\bar{V}_2$ are defined as $\bar{V}_1 = u - V_1$ and $\bar{V}_2 = v - V_2$. The effective active force along the $\mathbf{e}_1$ axis is given by $F_1^a = h m Q_{12} \, \partial_2 a_1 
+ \partial_2 \left[h a_1 m Q_{12}\right] 
+ \partial_1 \left[a_2 m h (Q_{11} - Q_{33})\right].$
The effective  active force parallel to the film along the $\mathbf{e}_2$ direction can be obtained by switching $1 \leftrightarrow 2$ in the derivatives.}\label{tablecurvedeq}
\end{table*}
where $F_1$ and $F_2$ are constants of integration, and  $\int^n$ denotes an indefinite integral. Integrating the next order transverse force balance across the thickness and using stress-free boundary conditions (Eq. \eqref{stressfree}), gives the effective perpendicular Stokes equation (see SM, Sec. \ref{forcebalanceeffective})
\begin{align}
& (2 \kappa_1+\kappa_2)(a_2 \partial_t a_1+\bar{V}_2 \partial_2 a_1+a_2 \partial_1 \bar{V}_1)\nonumber \\
& +\left(\kappa_1+2 \kappa_2\right)\left(a_1 \partial_t a_2+\bar{V}_1 \partial_1 a_2+a_1 \partial_2 \bar{V}_2\right) \nonumber \\
&+\frac{a_1 a_2 }{2 \mu }  m \left(\kappa_1 (Q_{11}-Q_{33})+\kappa_2 (Q_{22}-Q_{33})\right) =0.  \label{forcebalancentop}
\end{align}

We now turn to find the effective in-plane Stokes equations.
 Integrating the in-plane Stokes equation (Eq. \eqref{stokesequa}) between the two free surfaces, and applying the stress free boundary conditions, we find an effective Stokes equation along the $\textbf{e}_1$ and $\textbf{e}_2$ directions. These read
\begin{align}
& \partial_1 \{\frac{2\mu h }{a_{1} } [2\left(a_2 \partial_t a_1 +\bar{V}_2 \partial_2 a_1 +a_2 \partial_1\bar{V}_1\right) \nonumber \\
& +\left(a_1 \partial_t a_2+\bar{V}_1 \partial_1 a_2 +a_1  \partial_2\bar{V}_2\right)]\} +F_1^a\nonumber \\
& +\frac{1}{a_1} \partial_2\{\frac{\mu a_1  h }{a_2 }\left[a_1 \partial_2\bar{V}_1+a_2 \partial_1\bar{V}_2\right]\} \nonumber \\
& =\frac{2 \mu h }{a_1  a_2 } \partial_1 a_2 [\left(a_2 \partial_t a_1 +\bar{V}_2\partial_2 a_1 +a_2  \partial_1\bar{V}_1\right)\nonumber \\
& +2\left(a_1 \partial_t a_2 +\bar{V}_1 \partial_1 a_2+a_1 \partial_2\bar{V}_2 \right) ],\label{longforcebalance}
\end{align}
\begin{align}
&\partial_2\{\frac{ 2 \mu  h } {a_{2}} [\left( a_{2} \partial_t a_1+\bar{V}_2 \partial_2 a_1+ a_2 \partial_1\bar{V}_1\right)\nonumber \\
&+2\left(a_1 \partial_t a_2+\bar{V}_1 \partial_1 a_2+a_1 \partial_2 \bar{V}_2\right) ] \}+F_2^a \nonumber \\&
+\frac{\partial_1}{a_2} \left\{\frac{\mu a_2 h}{a_1}\left[a_1 \partial_2\bar{V}_1+a_2 \partial_1\bar{V}_2\right]\right\} \nonumber \\&
=\frac{2 \mu h}{a_1 a_2} \partial_2 a_1\left[2\left(a_2 \partial_t a_1+\bar{V}_2 \partial_2 a_1+a_2 \partial_1\bar{V}_1\right)\right. \nonumber \\&
\left.+\left(a_1 \partial_t a_2+\bar{V}_1 \partial_1 a_2+a_1 \partial_2\bar{V}_2\right)\right],\label{secondfbeq}
\end{align}
where the terms $F_1$ and  $F_2$ account for the effect of activity, and read
\begin{align}
&F_1^a=  h  m Q_{12} \partial_2 a_1 
+  \partial_2 [h a_1 m Q_{12}]+  \partial_1 [ a_2 m h(Q_{11}-Q_{33})],\\
&F_2^a=  h  m Q_{12} \partial_1 a_2
+  \partial_1 [h a_2 m Q_{12}]+  \partial_2 [ a_1 m h(Q_{22}-Q_{33})].
\label{activey} 
\end{align}
Eqs. \eqref{dynamicsofS}--\eqref{dynamicsoftheta}, ~\eqref{relat}--\eqref{gg3}, \eqref{masscon}, \eqref{forcebalancentop}, ~\eqref{longforcebalance}--\eqref{secondfbeq} govern the asymptotic dynamics of the curved film for the fields nematic orientation $\theta$, nematic order $S$, film thickness $h$, fluid velocity along the center-surface $u,v$, center-surface velocity of the film $V_1,V_2,V_3$,
and the geometric scaling factors $a_1,a_2$, and curvature along the curvature lines $\kappa_1$ and $\kappa_2$. These are summarized in Table \ref{tablecurvedeq}.\\

In the absence of activity ($m = 0$), the above equations reduces to the equations governing shape changes in curved, viscous thin fluid films. A pressure drop $\Delta P$ across the film can be readily incorporated as an external force into Eq.~\eqref{forcebalancentop}, which then becomes
\begin{align}
& (2 \kappa_1+\kappa_2)(a_2 \partial_t a_1+\bar{V}_2 \partial_2 a_1+a_2 \partial_1 \bar{V}_1)+ \nonumber \\
& +\left(\kappa_1+2 \kappa_2\right)\left(a_1 \partial_t a_2+\bar{V}_1 \partial_1 a_2+a_1 \partial_2 \bar{V}_2\right) =\frac{a_1 a_2 \Delta P }{2 \mu h}.  \label{forcenoac}
\end{align}
To gain further insight into Eq.~\eqref{forcenoac}, it is instructive to consider several limiting cases.
 For an axisymmetric bottle with radius $R$ {aligned} along z axis, we have
\begin{align}
a_{z}= 1, \quad a_{\theta}=R, \quad
\kappa_{z}=0,\quad \kappa_{\theta}=\frac{1}{R},
\end{align}
and Eqs. \eqref{masscon} and \eqref{forcenoac} simplify to \cite{van1995pressure}
\begin{align}
&\partial_t\left(R h\right)=0\label{massconsph}, \\
&  \partial_t R  =\frac{R^2 \Delta P}{4 h \mu} . 
\end{align}
For a constant pressure difference over time, these equations can be readily solved to yield ~\cite{howell1994extensional}
\[
R = R_0\left(1 - \frac{\Delta P R_0 t}{2 \mu h_0}\right)^{-\frac{1}{2}}, \quad 
h = h_0\left(1 - \frac{\Delta P R_0 t}{2 \mu h_0}\right)^{\frac{1}{2}},
\]
indicating that the radius increases while the thickness vanishes in finite time. However, in most biological settings, this singular behavior is likely to be ameliorated by by the role of activity, non-linear rheology etc.
 
For a sphere with radius $R$, the scaling factors and principal curvatures are given by
$$
\begin{gathered}
a_{\phi}= R \sin \theta, \quad a_{\theta}=R, \quad
\kappa_{\phi}=\kappa_{\theta}=\frac{1}{R}.
\end{gathered}
$$
Supposing spherical symmetry, the equations of motion in \eqref{masscon} and \eqref{forcenoac} may be written in the form \cite{van1995pressure}
\begin{align}
&\partial_t\left(R^2  h\right)=0,\label{massconsph} \\
&  \partial_t R  =\frac{R^2 \Delta P}{12 h \mu}. \label{masscong}
\end{align}
Assuming a constant pressure difference over time, these equations can be readily solved to yield~\cite{howell1994extensional} 
\[
R = R_0\left(1 - \frac{\Delta P R_0 t}{4 \mu h_0}\right)^{-\frac{1}{3}}, \quad 
h = h_0\left(1 - \frac{\Delta P R_0 t}{4 \mu h_0}\right)^{\frac{2}{3}}.
\]
{As in the cylindrical case, the radius blows up while the thickness vanishes in finite time; however, the scaling behavior is different.}

In Sec. \ref{sectionlsa}, we apply these equations to perform a linear stability analysis and examine the instabilities that arise in the nematic and isotropic phases of an active cylindrical thin film.

\subsection{Linear stability analysis of a cylindrical film}\label{sectionlsa} 
Since we incorporated the extrinsic properties into the formulation of the equations, the dynamics of the nematic order parameter yield the correct steady-state solution on a cylinder, which corresponds to the alignment of the nematic order along the long axis of the cylinder which we denote by $\textbf{e}_1$.

We define the dimensionless perturbation wavevector along the cylinder’s long axis as $\mathscr{Q}=q_x R$ and employ the Fourier representation $\delta f(q_x,\mathscr{P},\omega)=\int da_1 da_2 f(a_1,a_2) e^{-i(\mathscr{P} \theta+q_x x)+\omega t}$ for the perturbations in all fields. We then use Eqs. \eqref{relat}–\eqref{gg2} and \eqref{longforcebalance}- \eqref{secondfbeq} to solve for the steady-state values of the unknown fields and their perturbations $( \delta V_1, \delta V_2, \delta \kappa_1, \delta \kappa_2, \delta a_1, \delta a_2,\delta u, \delta v)$, which are then substituted into the dynamics of the nematic order parameter (Eqs. \eqref{dynamicsofS}-\eqref{dynamicsoftheta}), thickness (Eq. \eqref{masscon}), and the effective Stokes equations perpendicular to the surface (Eq. \eqref{gg2}), to find the dynamics of the perturbations in the orientation field $\delta \theta$, magnitude of the nematic order $\delta S$, thickness $\delta h$, and the center-surface $\delta V_3$.

For a cylinder with radius $R_0$, the steady-state solutions are $\textbf{u}=\textbf{V}=0$, $\kappa_1=\theta=0$, $a_1=1$, and $\kappa_2=1/a_2=1/R_0$.
\subsubsection{Instability of the ordered phase} 
\begin{figure}[H] 
    \centering
\includegraphics[width=0.45\textwidth]{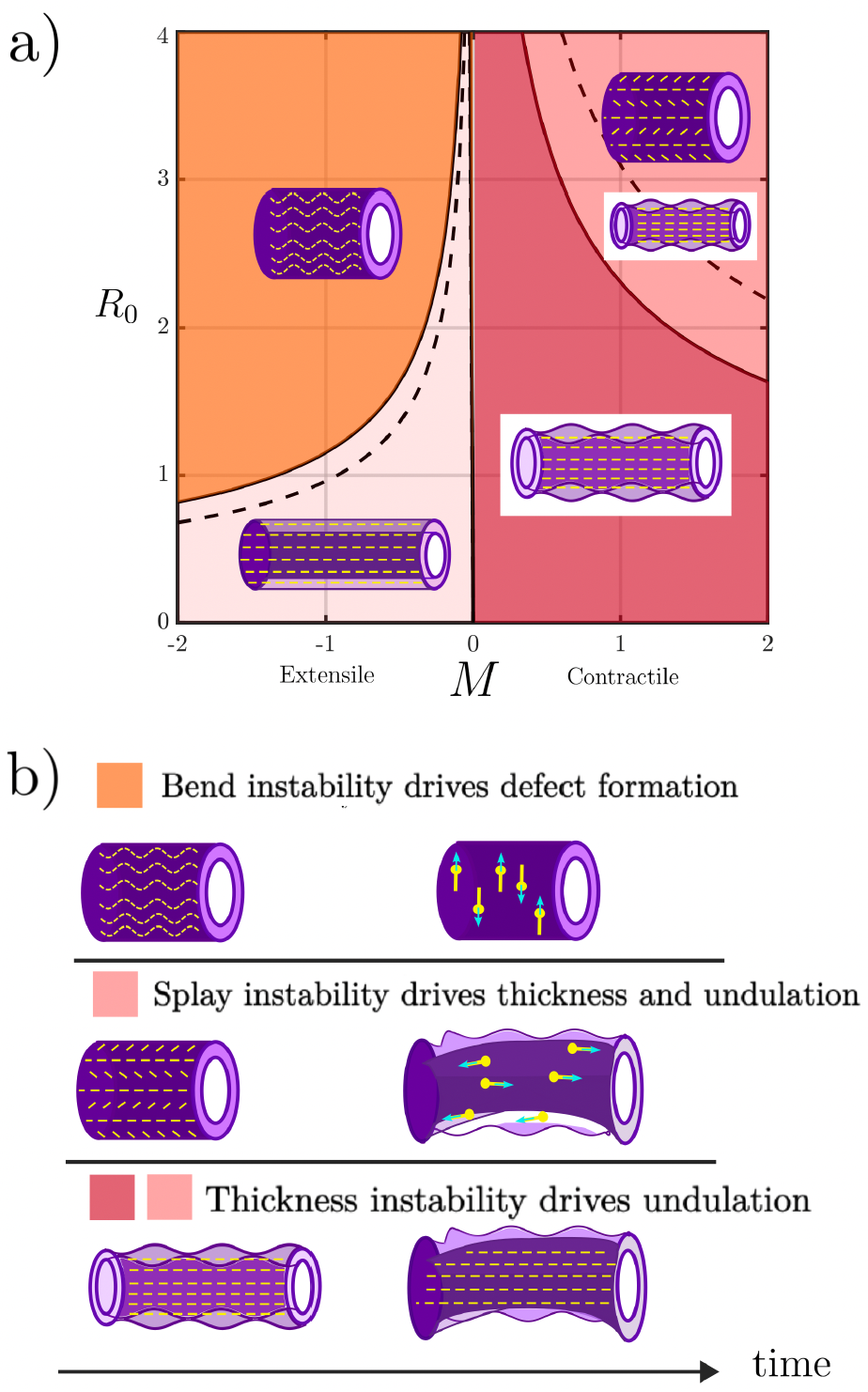}
    \caption{\textbf{Instability of the nematic phase of a living film with cylindrical geometry and finite thickness.} 
(a) Instability phase diagram as a function of the cylinder radius $R_0$ and rescaled activity $M = m \gamma /(\mu K_Q)$ in the nematic phase, as predicted by Eqs.~\eqref{nemcyl4}–\eqref{nemcyl6}. The phase boundaries are shown by solid (dashed) black lines for $\lambda = 0$ ($\lambda = 1$). Equation~\eqref{nemcyl4} predicts that for large radii in extensile systems ($M < 0$), bend perturbations grow and destabilize the nematic order (orange background), similar to what is observed in conventional 2D active nematics. At the linear level, this instability does not lead to shape changes or thickness modulations, suggesting that extensile activity does not alter the geometry of the cylinder. In addition, there exists a critical radius $R_c = \left(4 \mu K_Q / (-3 m \gamma)\right)^{1/2}$ below which the nematically ordered cylinder is stable (bright pink background). 
As predicted by Eq.~\eqref{nemcyl5}, contractile activity destabilizes the homogeneous thickness profile (red background), resulting in the emergence of thick and thin regions along the cylinder axis. Classical 2D active nematic theories, which neglect film thickness, cannot capture this thickness-driven instability. (b) The top row shows that the bend instability in the nematic field does not induce changes in thickness or cylinder shape, as the dynamics of thickness and shape (Eqs.~\eqref{nemcyl5} and \eqref{nemcyl6}) are decoupled from the nematic order dynamics. The middle and bottom rows illustrate that contractile activity leads to thickness and shape deformations of the cylinder, and to splay instabilities in the nematic field if the activity is sufficiently large.
 }\label{fig2b}
\end{figure}

As we explained for the linear stability analysis on the film, dynamics of the nematic tensor has a nematic steady state solution ($S=1$) for $\mathcal{B}=-2/3$. In this subsection, we study the stability of this phase. 
By substituting the Fourier decomposition of the fields ($\delta f(q_x,\mathscr{P},\omega$)) into equations \eqref{dynamicsofS}--\eqref{dynamicsoftheta}, ~\eqref{relat}--\eqref{gg3}, \eqref{masscon}, \eqref{forcebalancentop}, ~\eqref{longforcebalance}--\eqref{secondfbeq}, we find that, in the nematic phase, the dynamics of the perturbations in the nematic order parameter $\delta S$ do not lead to an instability. Specifically, the dynamics of the orientation of the order parameter and the thickness are decoupled from the center-surface, whereas the center-surface dynamics are coupled to both the thickness and the nematic order parameter. \\
Long-wavelength perturbations, associated with fluctuations along the cylinder's long axis ($\mathscr{P}=0$ and $\mathscr{Q} \to 0$), exhibit decoupled growth rates for the perturbations in the nematic director field $\omega_{\theta}$ and the thickness $\omega_{h}$, which satisfy the following equations:
\begin{align}
    &\omega_{\theta}=- \frac{K_Q}{2 R_0^2 \gamma}-\frac{m (9+4 \lambda)}{24 \mu}, \label{nemcyl4}\\
    &  \omega_{h} =\frac{m}{2 \mu}, \label{nemcyl5}\\
    & \delta  V_3= -\frac{\delta h m R_0 }{2 h_0 \mu}.\label{nemcyl6}
\end{align}
Eq. \eqref{nemcyl4} demonstrates that there exists a critical extensile activity   above which the ordered phase becomes unstable. This instability is characterized by the condition $-m \gamma /(\mu K_Q) > 4/(3 R_0^2)$, as depicted in the phase diagram in Fig. \ref{fig2b}(a). Notably, since in Eqs. \eqref{nemcyl5}-\eqref{nemcyl6} the perturbations in the thickness and center-surface ($w_h$ and $\delta V_3$) are decoupled from the orientation field $\delta S$
(Eq. \eqref{nemcyl4}), this instability does not induce any changes in thickness or shape, as illustrated in the time evolution shown in Fig. \ref{fig2b}(b), top row. {In addition,} Eq. \eqref{nemcyl5} demonstrates that contractile activity ($M > 0$) induces thickness instabilities, as illustrated in the phase 
diagram  in Fig. \ref{fig2b}(a). Since the center-surface perturbation is coupled to thickness perturbations through Eq. \eqref{nemcyl6}, the instability in thickness leads to shape deformation of the cylinder, as seen in the time evolution depicted 
in Fig. \ref{fig2b}(b), bottom. \\
For perturbations along the circumferential direction ($\mathscr{P} > 0$), we observe that
\begin{align}
&\omega_{\theta}=\frac{(9-4\lambda) m}{24  \mu}-\frac{c_1 K_Q}{R_0^2 \gamma},\label{eqomegathe}\\
&g_1 \delta h = -c_2 h_0 m \mathscr{Q} \delta \theta,\\
&\delta V_3=g_2 (\delta h, \delta \theta).\label{g2sm}
\end{align}
where $g_1=(c_3 \mathscr{Q}^2 m +(c_4 \mathscr{Q}^2 -1) \mu \omega)$, $c_{1}(\mathscr{P})>0$ is a scaler of order $\sim 1 - 10$ and, $c_2(\mathscr{P})>0$ and $c_3(\mathscr{P})>0$
are scalers of order $\sim 1/10 - 1$. The coefficient $c_{4}(\mathscr{P})\ge0$ is non-zero only for $\mathscr{P}=1$. 
The function $g_2$ takes different values depending on $\mathscr{P}$. The value of $g_2$ depends on 
$\mathscr{P}$, with two distinct values shown in SM, Sec. \ref{v3examples} for $\mathscr{P}=1$ and $\mathscr{P}=2$. The nature of the instability, which we discuss below, remains the same for various values of $\mathscr{P}>0$.\\
Equation \eqref{eqomegathe} defines the threshold contractile activity $ 3m /(8 \mu)\sim c_1 K_Q$ above which the nematic phase becomes unstable. This is illustrated in the phase diagram in Fig. \ref{fig2b}(a), top right. Since thickness and center-surface are coupled to the orientation field, once the instability initiates in the orientation field, it leads to instabilities in the thickness and shape of the cylinder, as shown in Fig. \ref{fig2b}(b), middle row.

The result of the linear stability analysis in this section shows that in addition to the well-studied instability of the director field in the nematic phase of the cylinder, which happens when the radius is larger than a critical radius \cite{pearce2020defect,al2023morphodynamics},  accounting for thickness changes, introduces a novel thickening instability (see Eq. \eqref{nemcyl5}) in contractile systems, that can happen even in cylinders with small radius (large curvature).

\subsubsection{Instability of the isotropic phase}
\begin{figure}[H] 
    \centering
\includegraphics[width=0.45\textwidth]{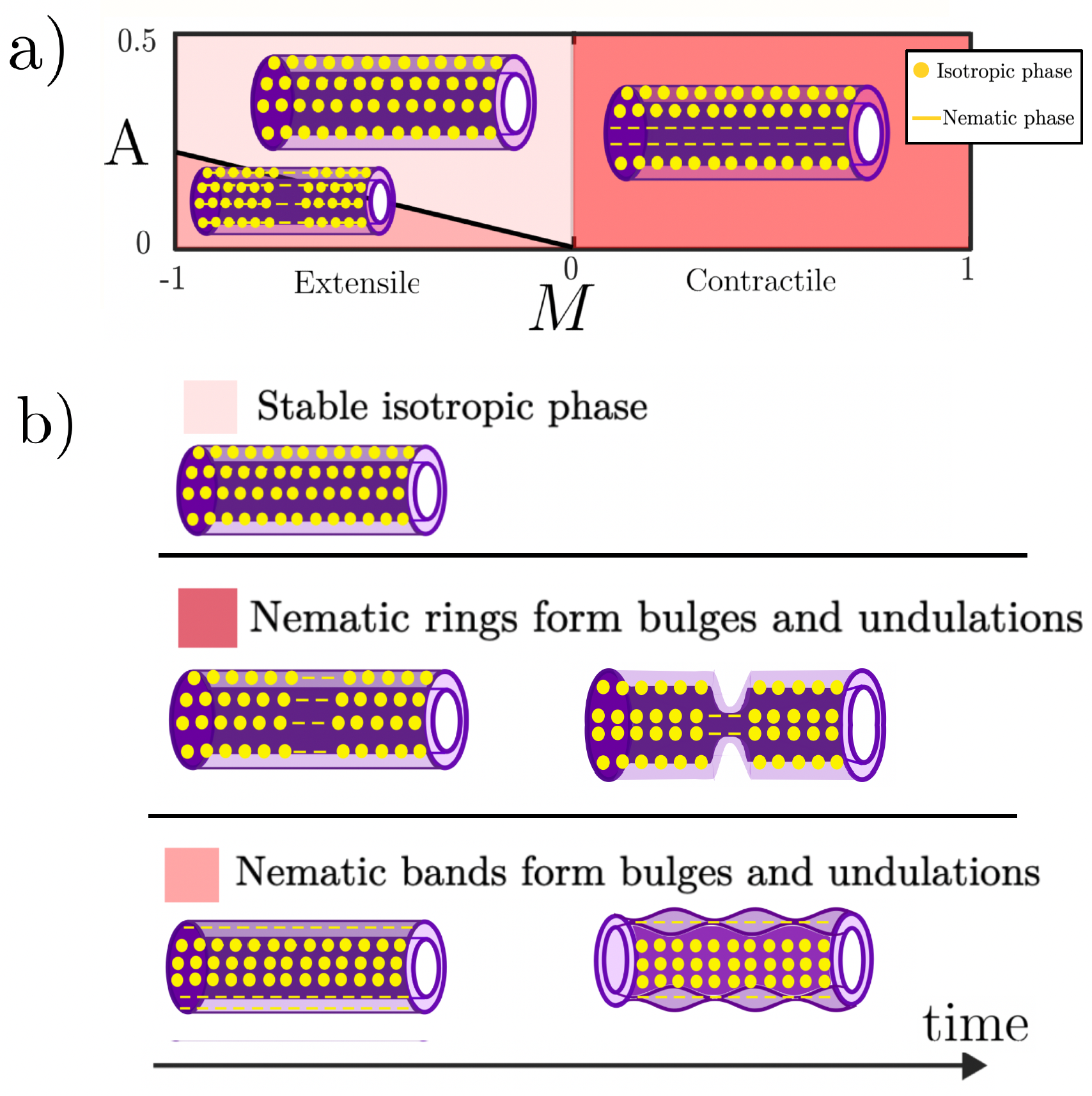}
    \caption{\textbf{Instability of the nematic phase of a living film with cylindrical geometry and finite thickness.}  
(a) Phase diagram of the instability of an initially isotropic phase (yellow disks), as predicted by Eqs.~\eqref{firstcyliso} and \eqref{activitysign}. The vertical axis shows the rescaled passive relaxation rate of the nematic phase back to the isotropic phase, $A = \mathcal{A} \mathcal{B} / K_Q$, determined by the free energy. Similar to 2D active nematics, sufficiently strong extensile activity ($|m|>9 \mathcal{A} \mathcal{B} \mu/(2 \gamma \lambda)$, pink region) induces nematic ordering (see Eq.~\eqref{firstcyliso}). In contrast to flat systems where the orientation of the emergent order is spontaneously selected, here the nematic order aligns along the cylinder’s long axis. Moreover, unlike conventional 2D active nematics on non-deformable surfaces—where contractile activity typically does not induce ordering—here, even contractile systems exhibit growth in perturbations of the nematic order magnitude along the circumferential direction (see Eq.~\eqref{activitysign}), resulting in the emergence of nematic order.
(b) Schematic representation of the instability described in panel (a) and its effects on other fields. Top row: For small extensile activity, the isotropic phase, uniform thickness, and cylindrical shape remain stable. Middle row: Sufficiently large extensile activity leads to the formation of {nematic order in} circumferential bands  which subsequently induce thickness modulations and deformations of the cylinder along the circumferential direction. Bottom row: Contractile activity leads to the formation of  {nematic order in bands  } aligned with the cylinder’s long axis, accompanied by corresponding changes in thickness and shape along the longitudinal direction.
 }\label{fig2}
\end{figure}

To study the stability of an isotropic phase, we use $\mathcal{B}=2/3$, which leads to a steady state solution ($S=0$). 
As before, substituting the Fourier decomposition of the fields, the growth rates of the perturbations in the nematic order parameter S, thickness h, 
and center-surface 
$V_3$, are respectively given by
\begin{align}
\omega_{S}= &- \gamma^{-1}(\frac{3 K_Q \left(\mathscr{Q}^2+\mathscr{P}^2\right)}{R_0^2}- \mathcal{A} \mathcal{B}) 
\nonumber\\&-\frac{  \lambda  m}{9  \mathscr{Q}^2 \mu }  \left(\mathscr{Q}^2-2 \mathscr{P}^2+2\right) \left( \mathscr{Q}^2+\mathscr{P}^2\right),
\label{cyliso}\\
&\delta h \omega = \frac{h_0 m  }{4 \mu}\delta S,\label{cylisog}\\
&\delta V_3=\frac{m  \left(2 \mathscr{P}^2-\mathscr{Q}^2\right)R}{4 \mathscr{Q} \mu  } \delta S. \label{cylisods}
\end{align}
Eqs. \eqref{cyliso} -\eqref{cylisods}, suggest that perturbations in the center-surface $\delta V_3$ and thickness $\delta h$ are coupled to the perturbations in the nematic order parameter $\delta S$. Therefore, 
any instability in the nematic order leads to instabilities in both the thickness and the shape. Equation \eqref{cyliso} indicates that the isotropic phase is more stable for $\mathscr{P}=1$ compared to $\mathscr{P}=0$. Thus, we focus on the growth rate for $\mathscr{P}=0$. For $\mathscr{P}=0$ and $\mathscr{Q}\to 0$,  the growth rate simplifies to
\begin{align}
\omega_{S}= - \mathcal{A} \mathcal{B}\gamma^{-1}
-\frac{ 2 \lambda  m}{9  \mu }.\label{firstcyliso}
\end{align}
Defining $A=\mathcal{A} \mathcal{B} /K_Q$, the phase diagram for the instabilities is shown in Fig. \ref{fig2} (a). For perturbations along the long axis where $\mathscr{P}=0$, the instability occurs in extensile systems when 
$A<-2 \lambda M/9$, resulting in the 
formation of loops of nematic order along the circumferential direction (bottom left phase in the phase diagram of Fig. \ref{fig2}(a)). This instability leads to changes in thickness, as illustrated in the schematic time dynamics shown in the middle panel of Fig. \ref{fig2}(b).

For large perturbations along the circumferential ($\mathscr{P}\geq2$) direction, the coefficient of the active term changes sign, thereby altering the nature of the instability, which now occurs exclusively in contractile systems. This instability is initiated by the formation of nematic ordering in bands along the cylinder's axis ($M>0$ in the phase diagram in Fig. \ref{fig2} (a)), and subsequently leads to changes in thickness and shape deformation (see Eqs.
\ref{cyliso} -\ref{cylisods}), as illustrated in the time evolution schematic in Fig. \ref{fig2} (b). 

To compare this with the instability of an isotropic phase on a non-deformable cylinder, we first note that in the deformable case, shear stress ($\textbf{E} \sim \boldsymbol{\nabla} (\textbf{u} - \textbf{V})$) introduces an additional term ($\textbf{V} \neq 0$) due to deformability.

For non-deformable cylinders, the quantities $a_i$, $k_i$, and $V_i$ are constant both spatially and temporally, yielding the following expression for the perturbation growth rate
\begin{align}
\omega_S=- (\frac{3 K_Q \left(\mathscr{Q}^2+\mathscr{P}^2\right)}{ R_0^2}+ \mathcal{A} \mathcal{B}) \gamma^{-1}
- \frac{\lambda  \mathscr{Q}^2 m \left(\mathscr{Q}^2+7 \mathscr{P}^2\right)}{6 \mu  \left(\mathscr{Q}^2+\mathscr{P}^2\right)^2}.\label{activitysign}
\end{align}
It is easy to see that the sign of the coefficient of the activity in Eq. \eqref{activitysign} remains unchanged for various choices of $(\mathscr{P},\mathscr{Q})$, and as such the isotropic phase can only become 
unstable in extensile systems ($m<0$). 
This behavior is analogous to the instability of the isotropic phase in 2D active nematics \cite{PRXLife.1.023008,PhysRevLett.129.258001}.
\section{Discussion}\label{conclusionsection}
\begin{table*}[t]
    \centering
    \renewcommand{\arraystretch}{1.5} 
    \begin{tabular}{|p{4cm}|p{12cm}|}  
        \hline
        \textbf{Geometry} & \textbf{Stability analysis of the nematic phase} \\ 
        \hline
        Fixed 2D Cylinder & Perturbations along the long axis (circumference) destabilize the nematic phase in extensile (contractile) systems. The threshold activity is determined by the elasticity of the nematogens and the cylinder's radius. \\ 
        \hline
        Fixed 2D Flat Surface & Bend (splay) perturbations destabilize the nematic phase in extensile (contractile) systems, with no threshold activity for instability. \\ 
        \hline
        Cylindrical Film & In addition to the instabilities in the orientation field seen in the fixed cylindrical case, contractile activity leads to thickening, which induces shape instabilities. \\ 
        \hline
        Flat Film & Perturbations in all directions lead to instability in the nematic order, with the maximum growth rate for bend instabilities in both extensile and contractile systems. The instability growth rate is influenced by both thickness and nematic orientation, with no shape deformations.\\ 
        \hline
    \end{tabular}
    \caption{Comparison of instabilities in nematic phases for a 2D fixed cylinder, a 2D fixed plane, a cylindrical film, and a flat film.}
    \label{tab:phase_comparison}
\end{table*}
\begin{table*}[t]
    \centering
    \renewcommand{\arraystretch}{1.5} 
    \begin{tabular}{|p{4cm}|p{12cm}|}  
        \hline
        \textbf{Geometry} & \textbf{Stability analysis of the isotropic phase} \\ 
        \hline
        Fixed 2D Cylinder & Only extensile activity leads to the formation of order. The threshold activity for instability is set by the free energy cost of order formation. \\ 
        \hline
        Fixed 2D Flat Surface & Only extensile activity induces nematic order. The threshold activity for instability is determined by the free energy cost of order formation. \\ 
        \hline
        Cylindrical Film & Perturbations along the long axis lead to order formation in extensile systems. Unlike non-deformable cylinders, contractile activity also induces nematic order for circumferential perturbations. Instabilities in the magnitude of the nematic order lead to instabilities in the film’s thickness and shape. \\ 
        \hline
        Flat Film & Unlike on a fixed 2D flat surface, instabilities in the magnitude of the nematic order induce thickness instabilities. \\ 
        \hline
    \end{tabular}
    \caption{Comparison of instabilities in isotropic phases for a 2D fixed cylinder, a 2D fixed plane, a cylindrical film, and a flat film.}
    \label{tab:phase_comparisonb}
\end{table*}
We have derived asymptotic equations governing the dynamics of an active, deformable thin films of varying geometries embedded in three dimensions. The resulting equations couple the dynamics of thickness and shape deformations with the orientation of the active agents  {and fluid velocity} within the film,  as well as the velocity of the film’s center-surface.\\

In theories for conventional 2D active nematic systems, the predictions of linear stability analysis are consistent with both fully nonlinear simulations and experimental observations which show the growth of bend instabilities in the nematic phase of extensile systems~\cite{PhysRevLett.89.058101}, as well as the emergence of nematic order driven by extensile activity in initially isotropic systems in experiments and simulations ~\cite{opathalage2019self,PRXLife.1.023008}. 

Our current theory allows us to build on this understanding of 2D active nematic theories, generalized to include shape deformations—both in thickness and of the film center-surface. Although we have derived the asymptotic equations for the nonlinear dynamics of thin   films with a general geometry, in the current study we have limited ourselves to studying the  {linear} stability of homogeneous phases on flat and cylindrical thin films. The results are summarized in Figs.~\ref{fig3}, \ref{fig2b}, and \ref{fig2} for the flat and cylindrical films, respectively. As detailed in tables~\ref{tab:phase_comparison} and \ref{tab:phase_comparisonb}, we found that curvature couples shape, thickness, and order, enabling shape changes of the film even when activity remains homogeneous across the thickness. Furthermore, in nearly flat films, both extensile and contractile activity lead to instability of the nematic phase. This instability occurs for perturbations in all directions relative to the direction of the nematic order. This contrasts with conventional 2D active nematic instabilities, in which bend (splay) perturbations grow in extensile (contractile) systems.

A natural next step is to numerically solve the governing equations for weakly curved films, to better understand the nonlinear dynamics of thickening, deformation, and defect evolution—an open challenge for future research.\\

\section*{Acknowledgements}
We acknowledge helpful discussions with Ioannis Hadjifrangiskou and Farzan Vafa, and thank the Simons Foundation and the Henri Seydoux Fund for partial financial support.
\section{Supplementary Material (SM)}
\subsection{Dynamics of the nematic tensor relative to moving coordinates}\label{dynamicsofq}
The time derivative $\bar{D}_t$ in Eq. \eqref{eqdyq} is defined as $(\bar{D}_t \textbf{Q})^{m\delta} = \partial_t (Q_{ij} \hat{e}_i \hat{e}_j):\hat{e}^m \hat{e}^{\delta} +a_{\delta} a_m ( -\bar{V}^{c} \nabla_c Q^{m \delta})$, where $a_i$ shows the scaling factor of the curvilinear coordinate along $x_i$ axis, and \( \mathbf{\bar{V}} = \mathbf{u} - \mathbf{V} \) shows the fluid velocity respect to the velocity of the coordinate $\textbf{V}$. 
In the evolving coordinate system, the $_{m \delta}$ component of the dynamics of the nematic tensor (Eq. \eqref{eqdyq}) is then written as:
\begin{align}
& \partial_t (Q_{ij} \hat{e}_i \hat{e}_j):\hat{e}_m \hat{e}_{\delta} = \partial_t Q_{m \delta} + Q_{\delta i} \hat{e}_m \cdot \partial_t \hat{e}_i + Q_{m i} \hat{e}_{\delta} \cdot \partial_t \hat{e}_i \nonumber \\
& = a_{\delta} a_m \Gamma (K \nabla^2 Q^{lm} - \mathcal{A}(1 + \mathcal{B} g_{\beta \gamma} g_{\alpha \delta} Q^{\alpha \beta} Q^{\gamma \delta}) \mathcal{B} Q^{ij}) \nonumber \\
& \quad + a_{\delta} a_m ( -V^{c} \nabla_c Q^{m \delta} + \frac{2}{d} \lambda E^{m \delta} + \Omega^m_\alpha Q^{\alpha \delta} - Q^{m \beta} \Omega_\beta^{\delta} ), \tag{S1} \label{eqmolecb}
\end{align}
where the strain rate $\textbf{E}$ and vorticity $\boldsymbol{\Omega}$ are computed using the velocities $\bar{\textbf{V}}$, which account for the evolution of the coordinate system.
\subsection{Laplacian of the nematic tensor}\label{nematicdiffusion}
We now define the Laplacian of contravariant nematic tensor used in Eq. \eqref{eqmolec} which reads
\begin{align}
\nabla^2 Q^{\mu \nu} =&  \nabla_k \nabla^k Q^{\mu \nu}= g^{\rho k} \nabla_k \nabla_\rho Q^{\mu \nu}= g^{\rho k} \nabla_k T^{\mu \nu}_\rho,\tag{S2}\label{lap}
\end{align}
where we have used the Christofel symbol $\boldsymbol{\Gamma}$ to define 
\begin{align}
T^{\mu \nu}_\rho= \nabla_\rho Q^{\mu \nu}= \partial_\rho Q^{\mu \nu}+\Gamma^\mu_{\alpha \rho} Q^{\alpha \nu}+\Gamma^\nu_{\alpha \rho} Q^{\mu \alpha}. \tag{S3}
\end{align}

Using the definition of $\textbf{T}$, Eq. \eqref{lap} then can be written as
\begin{align}
\nabla^2 Q^{\mu \nu} =&  g^{\rho k} (\partial_k T^{\mu \nu}_\rho+\Gamma^\mu_{k \alpha} T^{\alpha \nu}_\rho+\Gamma^\nu_{k \alpha} T^{\mu \alpha}_\rho-\Gamma^\alpha_{k \rho} T^{\mu \nu}_\alpha),\tag{S4}\label{lapb}
\end{align}
where
\begin{align}
\partial_k T^{\mu \nu}_\rho=& \partial_k \partial_\rho Q^{\mu \nu}+\partial_k \Gamma^\mu_{\alpha \rho} Q^{\alpha \nu}
+ \Gamma^\mu_{\alpha \rho} \partial_k Q^{\alpha \nu}+\partial_k \Gamma^\nu_{\alpha \rho} Q^{\mu \alpha} \nonumber\\&+ \Gamma^\nu_{\alpha \rho} \partial_k Q^{\mu \alpha}.\tag{S5}
\end{align}
\subsection{Liquid crystal dynamics}\label{sthetadynamics}
In this subsection, we show how to use 
the Pauli matrices $\boldsymbol{\Pi}_{q}$ to drive the dynamics of the nematic order parameter $S$ and the direction of the order $\theta$ in Eqs. \eqref{dynamicsofS}   and \eqref{dynamicsoftheta} from the dynamics of the nematic tensor in \eqref{eqdyq}. 
Using the definition of Pauli matrices, we obtain  
\begin{align}
(\boldsymbol{\Pi}_p \boldsymbol{\Pi}_p)_{ij} = (\boldsymbol{\pi} \boldsymbol{\pi})_{ij} = \delta_{ij} - \delta_{i3} \delta_{j3},\tag{S6}
\end{align}
which corresponds to the 3D identity matrix with the \((3,3)\) component subtracted.  
The time evolution of \(\boldsymbol{\Pi}_p\) is given by  
\begin{align}
\frac{d \boldsymbol{\Pi}_p}{dt} = 2 \boldsymbol{\pi} \frac{d \theta}{dt}.\tag{S7}
\end{align}
Consequently, the time evolution of the nematic tensor can be expressed as  
\begin{align}
\partial_t \boldsymbol{Q} &= \frac{3}{4} \left( \boldsymbol{\Pi}_p + \boldsymbol{G} \right) \frac{d S}{dt} + \frac{3 S}{4} \frac{d \boldsymbol{\Pi}_p}{dt} \nonumber \\
&= \frac{3}{4} \left( \boldsymbol{\Pi}_p + \boldsymbol{G} \right) \frac{d S}{dt} + \frac{3 S}{2} \boldsymbol{\pi} \frac{d \theta}{dt}.\tag{S8}
\end{align}
Multiplying Eq.~(1) on the left by \(\boldsymbol{\Pi}_p\), taking the trace, and using  
\begin{align}
\operatorname{Tr}(\boldsymbol{\Pi}_p \boldsymbol{\Pi}_p) = 2, \quad
\operatorname{Tr}(\boldsymbol{\Pi}_p \boldsymbol{G}) = \operatorname{Tr}(\boldsymbol{\Pi}_p \boldsymbol{\pi}) = 0,\tag{S9}
\end{align}
we obtain the dynamics of the nematic order in Eq. \eqref{dynamicsofS}. Similarly, multiplying Eq.~(1) on the left by \(\boldsymbol{\pi}\), taking the trace, and using  
\begin{align}
\operatorname{Tr}(\boldsymbol{\pi} \boldsymbol{\pi}) = 2, \quad
\operatorname{Tr}(\boldsymbol{\pi} \boldsymbol{G}) = \operatorname{Tr}(\boldsymbol{\pi} \boldsymbol{\Pi}_p) = 0,\tag{S10}
\end{align}
yields the dynamics of the orientation field in Eq. \eqref{dynamicsoftheta}.
\section{Stress free boundary conditions}\label{bcsection}
To find the stress free boundary condition at the interface, we define the unit vector normal to the interface as 
\begin{equation}
 \textbf{N}=\left|\frac{\partial \boldsymbol{r}}{\partial x_1} \times \frac{\partial \boldsymbol{r}}{\partial x_2}\right|^{-1}\left(\frac{\partial \boldsymbol{r}}{\partial x_1} \times \frac{\partial \boldsymbol{r}}{\partial x_2}\right),\tag{S11}
\end{equation}
where {$\boldsymbol{r}(x_1,x_2,t)= \textbf{r}_c(x_1,x_2,t)\pm \frac{h}{2} \textbf{n}(x_1,x_2,t)$}.
Substituting normal to the interface {$\textbf{N}(x_1,x_2,t)$} in the  stress free boundary conditions  $\boldsymbol{\sigma}\cdot\textbf{N}|_{n=\pm h/2}=0$, we have
\begin{align}
\sigma_{13} & = \pm \frac{1}{2 l_1} \frac{\partial h}{\partial x_1} \sigma_{11} \pm \frac{1}{2 l_2} \frac{\partial h}{\partial x_2} \sigma_{12}, \tag{S12}\label{bcs11}\\
\sigma_{23} & = \pm \frac{1}{2 l_1} \frac{\partial h}{\partial x_1} \sigma_{12} \pm \frac{1}{2 l_2} \frac{\partial h}{\partial x_2} \sigma_{22}, \tag{S13}\label{bcs22}\\
\sigma_{33} & = \pm \frac{1}{2 l_1} \frac{\partial h}{\partial x_1} \sigma_{13} \pm \frac{1}{2 l_2} \frac{\partial h}{\partial x_2} \sigma_{23},\tag{S14}\label{bcs33}
\end{align}
where the scaling factors are defined as
$l_1=a_1\left(1-\kappa_1 n\right)$, and
$ l_2=a_2\left(1-\kappa_2 n\right)$.
\subsection{Expansion of the stress tensor and the dynamics in a nearly flat film}\label{stressexpansion}  
In this subsection, we define the lowest order terms in the definition of the stress tensor and the Stokes equation. Using the scaling introduced in Sec. \ref{sectionflatlub}, we can write the following expansion for the stress tensor
\begin{align}
        &{\sigma}_{11}=2 \mu \partial_{1} {u}^{(0)}+m Q_{11}-P^{(0)}+\mathcal{O}(\epsilon),\tag{S15}\label{sad1}\\& 
    {\sigma}_{12}= \mu ( \partial_{1} {v}^{(0)}+ \partial_{2} {u}^{(0)})+m Q_{12}-P^{(0)}+\mathcal{O}(\epsilon),\tag{S16}\label{sad2}\\& 
{\sigma}_{22}=2\mu \partial_{2} {v}^{(0)}+m Q_{22}+\mathcal{O}(\epsilon),\tag{S17}\label{sad3}\\&
{\sigma}_{13}=\mu \epsilon^{-1} \partial_{n}  {u}^{(0)}+\mathcal{O}(\epsilon^0),\tag{S18}\label{sad4}\\& 
{\sigma}_{23}=\mu \epsilon^{-1} \partial_{n} {v}^{(0)}+\mathcal{O}(\epsilon^0),\tag{S19}\label{sad5}\\& 
{\sigma}_{33}=2\mu \partial_{n} {w}^{(0)} -P^{(0)}+m Q_{33}+\mathcal{O}(\epsilon),\tag{S20}\label{sad6}
\end{align}
The components of the Stokes equation (Eq. \eqref{stokesequa}), parallel to the plane, can then be written as 
\begin{align}
        & \mu(\epsilon^{-2} \partial_n^2 {u}^{(0)}+\partial_n^2 {u}^{(2)})+\mu (\partial_{1}^2+\partial_{2}^2) {u}^{(0)}-\partial_1 P^{(0)}\nonumber \\&+m (\partial_1 Q_{11}+\partial_2 Q_{12})+\mathcal{O}(\epsilon)=0,\tag{S21}\label{sadsad1}\\
        &\mu(\epsilon^{-2} \partial_n^2 {v}^{(0)}+\partial_n^2 {v}^{(2)})+\mu (\partial_{1}^2+\partial_{2}^2) {v}^{(0)}-\partial_2 P^{(0)}\nonumber \\&+m (\partial_2 Q_{22}+\partial_1 Q_{12})+\mathcal{O}(\epsilon)=0.\tag{S22}\label{sadsad2}
\end{align}
Stokes equation (Eq. \eqref{stokesequa}) perpendicular to the center surface can then be written as 
\begin{align}
        & \mu(\epsilon^{-1} \partial_n^2 {w}^{(0)}+\epsilon \partial_n^2 {w}^{(2)})+\epsilon \mu (\partial_{1}^2+\partial_{2}^2) {w}^{(0)}-\epsilon^{-1}\partial_n P^{(0)}\nonumber \\&-\epsilon \partial_n P^{(2)}+m (\partial_1 Q_{13}+\partial_2 Q_{23})+\mathcal{O}(\epsilon)=0.\tag{S23}\label{sadsadg1}
\end{align}
Finally, the {incompressibility} condition can be written as
\begin{align}
        & \partial_n {w}^{(0)} + \partial_1 {u}^{(0)} + \partial_2 {v}^{(0)} + \epsilon^2 (\partial_n {w}^{(2)} + \partial_1 {u}^{(2)} + \partial_2 {v}^{(2)})=0.\tag{S24}\label{incompnnn}
\end{align}
\subsection{Torque balance for a nearly flat film}\label{torquebalanceflat}
To find the torque balance governing the dynamics of the flat film, we need to use the first-order non-zero elements of the stress tensor. Inserting the solution for the zero-order pressure (Eq. \eqref{pressureflat}) into the stress tensor, the first-order non-zero elements of the stress tensor read
\begin{align}
    &{\sigma}_{11}^0=2 \mu (2 \partial_{1} {u}^{(0)}+\partial_{2} {v}^{(0)})+m (Q_{11}-Q_{33}),\tag{S25}\label{zerostress0}\\& 
    {\sigma}_{12}^0=2 \mu ( \partial_{1} {u}^{(0)}+2 \partial_{2} {v}^{(0)})+m Q_{12},\tag{S26}\label{zerostress1}\\& 
{\sigma}_{22}^0=\mu (\partial_{1} {u}^{(0)}+2\partial_{2} {v}^{(0)})+m (Q_{22}-Q_{33}).\tag{S27}\label{zerostress}
\end{align}
The torque balance on the elements of the film can then be written as 
\begin{align}
 &\partial_{1}^2\int_{n^-}^{n^{+}}  dz z {\sigma}_{11}^{(0)}+ \partial_{2}^2 \int_{n^-}^{n^{+}} dz z {\sigma}_{22}^{(0)}
       + 2 \partial_{12} \int_{n^-}^{n^{+}}  dz z {\sigma}_{12}^{(0)}=0,\tag{S28}\label{eq26b}
\end{align}
where $n^-={{H}_0-\frac{{h}_0}{2}}$, and $n^+={{H}_0+\frac{{h}_0}{2}}$.
Substituting the stress tensor from Eqs. \eqref{zerostress0}-\eqref{zerostress} into the torque balance Eq. \eqref{eq26b} and evaluating the integrals, we obtain Eq.~\eqref{torquebalance} in the manuscript.
\subsection{Dispersion relation of the nematic phase in a nearly  flat film}\label{nematicflat}
In Eq. \eqref{eqwe}, we showed the growth rate of perturbations in the nematic phase in the absence of the flow aligning parameter  {$\lambda$}. The full growth rate of the perturbations in the presence of the flow aligning parameter $\lambda$ reads
\begin{align}
&\omega=\frac{1}{192 \mu} (A_1\pm\sqrt{B_1})-\frac{3 K_Q q^2}{4 \gamma}(1\pm \frac{C_1}{\sqrt{B_1}}),\tag{S29}\label{fullw}\\&
A_1=-2 m (5 \lambda +3 \lambda  \cos (4 \phi)+9 \cos (2 \phi)-9),\nonumber \\&
B_1=2 m^2 (\lambda  (59 \lambda +108)+162 (5 \lambda +6) \cos (2 \phi) \nonumber\\&+2 m^2 (9 \lambda  (\lambda  \cos (8 \phi)+6 \cos (6 \phi))+1539))\nonumber\\&+3 (20 \lambda  (\lambda +3)+27) \cos (4 \phi)) ,\nonumber\\&
C_1=2m (9+5 \lambda+27 \cos (2\phi)+3 \lambda \cos(4\phi)),\nonumber
\end{align}
where $B_1>0$ independent of values for $\lambda$, $m$, and $\phi$, and as a result the imaginary part of   $\omega$ is zero, and there is no oscillations. Equation \eqref{fullw} simplifies to the Eqs. \eqref{eqwc} and \eqref{eqwe} in the manuscript for $\lambda=0$.
\subsection{Effective Stokes equation and mass conservation in curved films}\label{forcebalanceeffective}
In this subsection, we find the leading order effective stokes equation and mass conservation in curved films.
Noting that the metric tensor is diagonal, the total stress tensor, which includes viscous stress, pressure, and active stress can be expressed as  
\begin{align}
&\sigma^{ij} = \mu \left( g^{ii} [\partial_i u^j + u^m \Gamma_{i m}^j] + g^{jj} [\partial_j u^i + u^m \Gamma_{j m}^i] \right) -P \delta^{ij}\nonumber \\
&= \mu \left( \ell_i^{-2} [\partial_i u^j + u^m \Gamma_{i m}^j] + \ell_j^{-2} [\partial_j u^i + u^m \Gamma_{j m}^i] \right)-P \delta^{ij},\tag{S30}
\end{align}
where the scaling factors are defined as
$l_1=a_1\left(1-\kappa_1 n\right)$, and
$ l_2=a_2\left(1-\kappa_2 n\right)$, and $\kappa_1$ and $\kappa_2$ are principal curvatures of the center-surface. 
The stress components to the order $\mathcal{O}(\epsilon)$ are given by  
\begin{align}
{\sigma}_{11}  &=-P^{(0)}+\frac{2 \mu}{a_1^{(0)}}(\frac{\partial {u}^{(0)}}{\partial x_1}+\frac{{v}^{(0)}}{a_2^{(0)}} \frac{\partial a_1^{(0)}}{\partial x_2}-a_1^{(0)} \kappa_1^{(0)} V_3^{(0)})\nonumber \\&+  m Q_{11}+\mathcal{O}(\epsilon), \tag{S31}\\
{\sigma}_{22}  &=-P^{(0)}+\frac{2 \mu}{a_2^{(0)}}(\frac{\partial {v}^{(0)}}{\partial x_2}+\frac{{u}^{(0)}}{a_1^{(0)}} \frac{\partial a_2^{(0)}}{\partial x_1}-a_2^{(0)} \kappa_2^{(0)} V_3^{(0)})\nonumber \\&+  m Q_{22}+\mathcal{O}(\epsilon),\tag{S32}\\
{\sigma}_{33} &=-P^{(0)}+2 \mu (\frac{\partial {w}^{(1)}}{\partial n}+ \frac{\partial {w}^{(0)}}{\partial n} \epsilon^{-1})\nonumber \\&+ m Q_{33}^{(0)}+\mathcal{O}(\epsilon)\tag{S33}\\
 {\sigma}_{12} &= \frac{\mu}{a_1^{(0)} a_2^{(0)}}(a_1^{(0)} \frac{\partial {u}^{(0)}}{\partial x_2}-{u}^{(0)} \frac{\partial a_1^{(0)}}{\partial x_2}+a_2^{(0)} \frac{\partial {v}^{(0)}}{\partial x_1})\nonumber \\&-{v}^{(0)} \frac{\partial a_2^{(0)}}{\partial x_1}+ m Q_{12}+\mathcal{O}(\epsilon), \tag{S34}\\
 {\sigma}_{13} &= \mu(\frac{\partial {u}^{(1)}}{\partial n}+\frac{\partial {u}^{(0)}}{\partial n} \epsilon^{-1}+\kappa_1^{(0)} {u}^{(0)}+\frac{1}{a_1^{(0)}} \frac{\partial {w}_0}{\partial x_1})\nonumber \\&+ m Q_{13}+\mathcal{O}(\epsilon), \tag{S35} \\
 {\sigma}_{23} &= \mu(\frac{\partial {v}^{(0)} }{\partial n}\epsilon^{-1}+\frac{\partial {v}^{(1)}}{\partial n}+\kappa_2^{(0)} {v}^{(0)}+\frac{1}{a_2^{(0)}} \frac{\partial V_3^{(0)}}{\partial x_2})\nonumber \\&+ m Q_{23}+\mathcal{O}(\epsilon),\tag{S36}
\end{align}
 {and the} incompressibility condition (Eq.~\eqref{incomp}) can be expressed as 
\begin{align}
\frac{\partial}{\partial x_1} \left(l_2 u\right) + \frac{\partial}{\partial x_2} \left(l_1 v\right) + \frac{\partial}{\partial n} \left(l_1 l_2 w\right) = 0. \tag{S37}\label{incomcur}
\end{align}
 The incompressibility equation \eqref{incomcur} to the leading order reads
\begin{align}
\frac{\partial {w}^{(1)}}{\partial n} = &   
 -\frac{1}{a_1^{(0)} a_2^{(0)}} (\frac{\partial}{\partial x_1} (a_2^{(0)} {u}^{(0)}) + \frac{\partial}{\partial x_2} (a_1^{(0)} {v}^{(0)}) )\nonumber \\
&+(\kappa_1^{(0)}+\kappa_2^{(0)}) V_3^{(0)},\tag{S38}
\end{align}
 {and is} subject to the $\mathcal{O}(\epsilon)$ kinematic condition at $n = \pm \frac{1}{2} h^{(0)}$  {which reads}
\begin{align}
 &{w}^{(1)}-\boldsymbol{n}^{(1)} \cdot \frac{\partial \boldsymbol{r}_c^{(0)}}{\partial t}-\boldsymbol{n}^{(0)} \cdot \frac{\partial \boldsymbol{r}_c^{(1)}}{\partial t}= \pm \frac{1}{2} \frac{\partial h^{(0)}}{\partial t} \nonumber\\&\pm \frac{1}{2 a_1^{(0)}} \frac{\partial h^{(0)}}{\partial x_1}({u}^{(0)}-V_1^{(0)})  \pm \frac{1}{2 a_2^{(0)}} \frac{\partial h^{(0)}}{\partial x_2}({v}^{(0)}-V_2^{(0)}). \tag{S39} \label{w1kin}
\end{align}  
These expressions give an equation for ${w}^{(1)}$ 
\begin{align}
{w}^{(1)} &= \boldsymbol{n}^{(1)} \cdot \frac{\partial \boldsymbol{r}_c^{(0)}}{\partial t} + \boldsymbol{n}^{(0)} \cdot \frac{\partial \boldsymbol{r}_c^{(1)}}{\partial t} 
+ \frac{n}{a_1^{(0)} a_2^{(0)}} [a_1^{(0)} a_2^{(0)} (\kappa_1^{(0)} \nonumber \\ 
&\quad + \kappa_2^{(0)}) V_3^{(0)} - \frac{\partial}{\partial x_1} (a_2^{(0)} {u}^{(0)}) - \frac{\partial}{\partial x_2} (a_1^{(0)} {v}^{(0)}) ], \tag{S40}\label{wveloc}
\end{align}
and mass conservation (Eq.~\eqref{masscon} in the manuscript). 
The Stokes equation  {(Eq. \eqref{stokesequa})} in the transverse direction, combined with the stress-free boundary condition (Eq.~\eqref{bcs33}), yields the leading-order pressure:  
\begin{align}
P^{(0)} &= \frac{2 \mu}{a_1^{(0)} a_2^{(0)}} [a_1^{(0)} a_2^{(0)} (\kappa_1^{(0)} + \kappa_2^{(0)}) V_3^{(0)} \nonumber \\ 
&\quad - \partial_1 (a_2^{(0)} {u}^{(0)}) - \partial_2 (a_1^{(0)} {v}^{(0)}) ] + m Q_{33}. \tag{S41}\label{pres}
\end{align}
Substituting the leading-order velocity expressions (Eqs.~\eqref{vel1delta} and \eqref{vel2delta}) into Eq.~\eqref{pres}, we find that the nonzero stress components, \(\sigma_{13}\), \(\sigma_{23}\), and \(\sigma_{33}\), are of order \(\mathcal{O}(\epsilon)\). 
As a result, the $\mathcal{O}(\epsilon^0)$ terms in the transverse Stokes equation satisfy 
\begin{align}
\frac{\partial \sigma_{33}^{(1)}}{\partial n} + \kappa_1^{(0)} \sigma_{11}^{(0)} + \kappa_2^{(0)} \sigma_{22}^{(0)} = 0. \tag{S42}\label{stktr}
\end{align}  
Using the third component of the stress-free boundary condition,  
$$\sigma_{33}^{(1)} = 0 \quad \text{on} \quad n = \pm \frac{1}{2} h^{(0)},$$  
we can integrate Eq. \eqref{stktr} across the thickness and obtain the effective transverse stress balance, as given in Eq.~\eqref{forcebalancentop}.
\subsection{Examples of \(\delta V_3\) for $\mathscr{P} \neq 0$ in the ordered phase}\label{v3examples}
We mentioned after Eq. \eqref{g2sm} that the nature of the instability remains the same for various values of $\mathscr{P}>0$.
As an example, for $\mathscr{P} = 1$, we have  
\begin{align}
\delta V_3 = -\frac{m R \omega \left(3 h_0 \mathscr{Q} \delta \theta + \delta h \left(\mathscr{Q}^2 - 2\right)\right)}{4 h_0 \mathscr{Q}^2 \mu \omega + h_0 m \left(2 \mathscr{Q}^4 + 9 \mathscr{Q}^2 + 4\right)}.\tag{S43}
\end{align}
For $\mathscr{P} = 2$, we have  
\begin{align}
\delta V_3 = & -\frac{\mathscr{Q}^2 m R^3 \omega \left(6 h_0 \mathscr{Q} \delta \theta + \delta h \left(\mathscr{Q}^2 - 8\right)\right)}{h_0 D}, \nonumber \\
D = & 256 \lambda K_Q + \mathscr{Q}^2 \left(16 \lambda K_Q \left(\mathscr{Q}^2 + 8\right) + 4 \mathscr{Q}^2 \mu R^2 \omega \right. \nonumber \\
& \left. + m R^2 \left(2 \mathscr{Q}^4 + 39 \mathscr{Q}^2 + 40\right)\right).\tag{S44}
\end{align}
\bibliographystyle{apsrev4-1}
\bibliography{references}
\end{document}